\documentclass[a4paper,11pt]{article}
\pdfoutput=1 

\usepackage{jheppub} 
\usepackage[utf8]{inputenc}
\usepackage{xcolor}
\usepackage{epstopdf}
\usepackage{graphicx}
\usepackage{epsfig}
\usepackage{dcolumn}  
\usepackage{bm}    
 \usepackage{caption}
 \usepackage{subcaption}
\usepackage{amssymb} 
\usepackage{amsmath,bm}
\usepackage{amsfonts}  
\usepackage{amsmath}  
\usepackage{slashed}  
\usepackage{enumitem}
\usepackage[mathscr]{euscript}
\usepackage{tabu}
\usepackage{epsfig}
   
\newcommand{\be}{\begin{equation}}
\newcommand{\ee}{\end{equation}}
\newcommand{\ba}{\begin{array}}
\newcommand{\ea}{\end{array}}
\newcommand{\bea}{\begin{eqnarray}}
\newcommand{\eea}{\end{eqnarray}}
\newcommand{\p}{\partial}
\newcommand{\wt}{\widetilde}
\newcommand{\nn}{\nonumber}

\newcommand{\g}{\gamma}
\newcommand{\m}{\mu}
\newcommand{\n}{\nu}

\usepackage{pgfgantt}
\usepackage{xcolor}
\usepackage[utf8]{inputenc}

\definecolor{barblue}{RGB}{153,204,254}
\definecolor{groupblue}{RGB}{51,102,254}
\definecolor{linkred}{RGB}{165,0,33} 

\usepackage{xcolor,colortbl}

 \title{\textbf{\textsf{Discontinuities of free theories on $AdS_2$
}}}
  \author{Justin R. David${}^{a}$, Edi Gava${}^{b,d}$, Rajesh Kumar Gupta${}^c$, K.S. Narain${}^d$}
\affiliation{ 
\vspace{.1cm} ${}^a$CHEP, Indian Institute of Science,
C.V. Raman Avenue, Bangalore 560012, India.\\
${}^{b}$INFN, sezione di Trieste, Italy.\\
${}^{c}$Department of Physics, Indian Institute of Technology Ropar, Rupnagar, Punjab 140001, India.\\
$^{d}$ICTP, Strada Costiera 11, 34151 Trieste, Italy.}
\emailAdd{justin@iisc.ac.in, gava@ictp.it, rajesh.gupta@iitrpr.ac.in, narain@ictp.it}

\abstract{
The partition functions  of free bosons as well as fermions on $AdS_2$   are not smooth  as a function of their masses. 
For free bosons, the partition function on $AdS_2$ is not smooth  when the mass saturates the Breitenlohner-Freedman bound. 
We show that the expectation value of the scalar bilinear on $AdS_2$ exhibits a kink at the BF bound and the change 
in slope of the expectation value  with respect to the mass is proportional to the inverse radius of $AdS_2$. 
For free fermions, when the mass vanishes the partition function exhibits a kink. 
We show that  
expectation value of the fermion bilinear  is  discontinuous and the jump in the expectation value
is proportional to the  inverse radius of $AdS_2$. 
We then show the  supersymmetric actions of  the chiral multiplet on $AdS_2\times S^1$ and the hypermultiplet on 
$AdS_2\times S^2$   demonstrate these features. 
The supersymmetric backgrounds are such that  as the ratio of the radius of $AdS_2$ to  $S^1$ or $S^2$ is  dialled, 
the partition functions as well as expectation of bilinears are not smooth   for each Kaluza-Klein mode on $S^1$ or $S^2$. 
Our observation is relevant for evaluating one-loop partition function in the near horizon geometry of 
extremal black holes. 
}
 
\begin{document}
 \maketitle
\flushbottom
 
 \section{Introduction}
 
 Near horizon geometries of extremal black holes, or BPS black holes in supersymmetric theories contain an Euclidean 
 $AdS_2$ which describes the geometry of the non-compact space, see for example in  \cite{Sen:2008vm}. 
Even near-extremal limit of a large class of black holes are described by a $AdS_2$ throat region, see for example in 
\cite{Kunduri:2007vf}. 
For both these situations it is important to study fields in the background of $AdS_2$ geometry. 
For instance in \cite{Mann:1997hm,Sen:2008vm,Banerjee:2010qc,Banerjee:2011jp,Sen:2012kpz,Sen:2012cj,Keeler:2014bra} 
the logarithmic corrections to black hole entropy were obtained by evaluating the 
partition function of various fields in the background of $AdS_2\times S^2$ which is the near horizon geometry of 
both BPS as well as extremal but non-supersymmetry black holes in $4$ dimensions. 

It has been argued that 
 the path integral in the  near horizon geometry of BPS black holes  in supersymmetric theories can be evaluated 
 exactly using the method of supersymmetric localization \cite{Banerjee:2009af,Dabholkar:2010uh,Dabholkar:2011ec}. 
 These techniques also involve evaluating one loop partition functions in the $AdS_2$ background. 
 It was pointed out in \cite{David:2018pex,David:2019ocd} 
 that there are obstructions to the  use of localization to evaluate partition function and  has been recently emphasised 
 in \cite{Sen:2023dps}.  This obstruction arises due to the fact that Killing spinor grows exponentially in the radial direction of $AdS_2$ and 
 maps normalizable modes to non-normalizable modes \footnote{$AdS_2$ with a non-trivial monopole background 
 for the $R$ symmetric admits constant Killing spinors \cite{Pittelli:2018rpl}.  However the $AdS_2$ near horizon geometries of supersymmetric black holes doe not contain such monopole backgrounds.}. 
 Therefore the  supersymmetric algebra does not close in the 
 space of normalizable modes. 
  
 In this paper we would like to point out another phenomenon  exhibited by  partition functions of  theories on   $AdS_2$.
 The 
 partition function of free bosons and fermions on $AdS_2$ are not smooth as a function of their masses. 
 This phenomenon is of course also seen in partition function of  free bosons or free fermions on $\mathbb{R}^2$ at the 
 massless limit. However in the case of $AdS_2$, we will see that physical observables like expectation value of
 bilinears have discontinuities which are determined by the size of $AdS_2$. 
We will then demonstrate that this phenomenon is present in 
supersymmetric actions which  are   obtained by Kaluza-Klein reduction on $AdS_2\times S^1$ or $AdS_2\times S^2$.
These actions 
 contain towers of Kaluza-Klein masses which can be dialled on changing the ratio of the radii of $AdS_2$ and the compact space. 
Therefore the resultant partition functions are not smooth as a function of this ratio.

 Consider the free boson  $\phi$  on $AdS_2$ with mass given by 
 \begin{equation}\label{mass1}
 m^2 = -\frac{1}{4L^2} + \frac{x^2}{L^2}\, ,
 \end{equation}
 where $L$ is the radius of $AdS_2$ and $x$ is a parameter that 
 can be dialled.  The partition function is not smooth at $x=0$, which is the point at which the mass 
 saturates the Breitenlohner-Freedman bound \cite{Breitenlohner:1982bm,Breitenlohner:1982jf}. 
 We show that  scalar bilinear  $\langle \phi^2\rangle $ has a kink at $x=0$ and the change in slope of the 
 expectation value at $x=0$ is proportional to $1/L$. 
 We first study  this phenomenon numerically and then  use the 
 Green's function of the boson on $AdS_2$ to analytically demonstrate the presence of the kink. 
 Studying the Fourier decomposition of the Green's function  along the angular direction of 
$AdS_2$ in
 more detail, allows us to  isolate the source of the kink  to the fact that the  normalisable wave functions 
 changes according across $x=0$. 
 
For free Fermions  on $AdS_2$ we consider the mass given by 
\begin{equation} \label{mass2}
m = \frac{x}{L}\, .
\end{equation}
We show that the partition function has a kink at $x=0$, which is the
Breitenlohner-Freedman bound for fermions in $AdS$ \cite{Amsel:2008iz,Dias:2019fmz}.  
We then evaluate the expectation value of the fermion bilinear and show that it has a discontinuity at $x=0$. 
The jump in the value of the expectation value is proportional to $1/L$. 
We demonstrate this discontinuity both by studying the partition function using numerics as well as analytically by 
examining the Green's function of the fermion in $AdS_2$.

As we emphasised earlier  supersymmetric   field theories on  $AdS_2$ play important role in evaluating 
one loop corrections to black hole entropy. In this paper we consider the simplest examples of supersymmetric theories that 
arise on considering near horizon geometries of BPS black holes. 
We first examine the case of the chiral multiplet on $AdS_2\times S^1$, let the radius of $S^1$ be $U$. 
This theory was considered in \cite{David:2016onq,David:2018pex,David:2019ocd} where 
the partition function in the presence of a background vector multiplet was evaluated using the Green's function 
method.  To keep the discussion simple we set the background vector multiplet to zero and evaluate the partition function. 
We show that the supersymmetric backgrounds are such that the parameter 
$x$ in (\ref{mass1}) and (\ref{mass2}) 
in the masses of the bosons and fermions depends on the Kaluza-Klein mode number on $S^1$ and the 
ratio of the radii, $L/U$. 
As this ratio is dialled we see that the partition function is not smooth when the parameter $x$ vanishes. 
The critical value of the ratio $L/U$ for which $x$ vanishes differs for each Kaluza-Klein mode.
It is also distinct for bosons and fermions. This implies that the partition function 
has countably infinite points at which it is not smooth or at which the 
expectation values of scalar or fermion bilinears behave anomalously. 

Finally we study the case of the hypermultiplet on $AdS_2\times S^2$.  This geometry  occurs in all near horizon 
geometries of BPS supersymmetric black holes in $4$ dimensions. 
The supersymmetric background we consider on   $AdS_2\times S^2$ involves magnetic monopole on the $S^2$, expectation 
value of the background vector multiplet scalar as well as the auxillary scalar. 
We show that for each angular momentum mode on $S^2$, the parameter $x$ depends on the ratio of the radii  of 
$AdS_2$ to $S^2$, 
$L/U$.  For the bosons of the hypermultiplet the  parameter $x$ in (\ref{mass1})  which determines their mass 
is such that it does not vanish as the ratio is dialled for any angular momentum mode. 
However, for the fermions we show there exits a ratio $L/U$ for each angular momentum mode at which 
$x$ vanishes. Therefore, the partition function is not smooth and the fermion bilinear is discontinuous at these countably 
infinite set of points.  

The paper is organised as follows. 
In section \ref{boson} and section \ref{fermion}
 we study the partition function of bosons and fermions on $AdS_2$ as the function of 
their masses. In section \ref{localactions} we show that supersymmetric actions of the chiral multiplet on $AdS_2\times S^1$
and the hypermultiplet on $AdS_2\times S^2$ are not smooth as the function of the ratio of the radii of $AdS_2$ and the 
compact space. 
The appendix \ref{appen}  contains the details of the conventions and notations used in the paper. 
It also introduces monopole harmonics which is used to evaluate the partition functions on $AdS_2\times S^2$.

 \section{Bosons on $AdS_2$} \label{boson}
 
 In this section we study the free boson partition function and the scalar bilinear as a function of the mass. 
 We first perform the study numerically and then in section \ref{sec-greens} we evaluate  the expectation value of scalar bilinear 
 using the Green's function on $AdS_2$ and taking the coincident limit. 
 Finally  in \ref{sec-fourier}
 we demonstrate that the presence of kink in the scalar expectation value due to the fact that there is a change 
 in the normalizable wave function as the mass is varied. This is done using the Fourier decomposition of the Green's function.

 \subsection{The partition function  and the kink in $\langle \phi^2 \rangle$: numerics}
 \label{sec1.1}
 
 Consider the partition function of the free boson on $AdS_2$ with the following mass squared
 \begin{equation} \label{frebmas}
 m^2 = -\frac{1}{4 L^2} +   \frac{x^2}{L^2}\, .
 \end{equation}
 Here $L$ is the radius of $AdS_2$. We will study the the free energy as a function of   $x$, which parametrizes the 
 deviation of the mass from the Breitenlohner-Freedman bound.. 
 We will demonstrate that the partition function is not a smooth function in $x$   at the
 Breitenlohner-Freedman bound, that is when $x=0$.  We parametrize the deviation of the mass from the Breitenlohner-Freedman 
 by $x^2$ rather than say a linearly by setting $x^2 =y$ because we wish to reproduce the Localizing action
 of the boson on $AdS_2\times S^1$ as in (\ref{mbosact}) and examine how the partition function behaves on changing 
 the ratio of the radius of $AdS_2$ to $S^1$
  \footnote{If we were to parametrize the mass as 
 $ m^2 = -\frac{1}{4 L^2} +   \frac{y}{L^2}$ with $y>0$, the second derivative of the partition function with respect to $y$ is singular at 
 $y=0$.}.
 The action of the theory is given by 
 \begin{equation} \label{adsaction}
 S= \frac{1}{2} \int d^2x \sqrt{g} ( g^{\mu\nu} \partial_\mu \phi  \partial_\mu \phi  +  m^2 \phi^2 ) \, , 
 \end{equation}
 where the metric on $AdS_2$ is given by 
 \begin{equation} \label{metads2}
 ds^2 = L^2 ( dr^2 + \sinh^ 2 r d\theta^2) \, .
 \end{equation}
 The free energy of  the scalar is given  by 
 \begin{equation}
 -\log Z= \frac{1}{2} {\rm Tr} \log\Big( - \nabla  +   m^2\big ) \, , 
 \end{equation}
 where $-\nabla$ is the scalar Laplacian on $AdS_2$. It is know that the Laplacian 
  admits $\delta$ function normalizable eigen functions which are given by  \cite{Camporesi:1994ga}
  \begin{eqnarray} \label{scalwave}
  f_{\lambda\, p} (r, \theta)  &=& \frac{1}{\sqrt{2\pi L^2}}  \frac{1}{2^{|p| }    |p|! } 
  \left| \frac{\Gamma ( i \lambda  + \frac{1}{2} + |p|) }{ \Gamma( i \lambda) } \right| e^{i p \theta} 
  \sinh^{|p|} r \\ \nonumber
  && \times {}_2 F_{1} (  i \lambda + \frac{1}{2}  +|p| ,  -i \lambda + \frac{1}{2} + |p|, |k| + 1, - \sinh^2 \frac{r}{2} ) , 
  \\ \nonumber
  & & p \in \mathbb{Z}, \quad 0 \leq \lambda <\infty \, .
  \end{eqnarray}
  These functions satisfy
  \begin{equation}
  -\nabla   f_{\lambda\, p} (r, \theta) = \frac{1}{L^2} ( \frac{1}{4}  +\lambda^2) \, .
  \end{equation}
 To evaluate the partition function, it is convenient to examine  the heat kernel 
 \begin{equation}
 K ( r, \theta,  r', \theta'; t) =
  \sum_{p}\int d\lambda e^{ - t ( -\nabla + m^2) } f_{\lambda\, p} (r, \theta) f_{\lambda\, p}^* (r,' \theta')\, .
 \end{equation}
 For the partition function, we need to take the coincident limit and since $AdS_2$ is a homogenous 
 space, we can take this point to be the origin. 
 Now, the wave functions in (\ref{scalwave})  vanish at the origin for all $p\neq 0$, therefore it is 
 sufficient to look at the $p=0$ wave function, which is 
 \begin{equation}
 f_{\lambda\, 0} ( \eta, \theta) = \frac{1}{\sqrt{ 2\pi L^2} }\sqrt{ \lambda \tanh (\pi \lambda) } \, .
 \end{equation}
 Substituting this for the coincident limit, we obtain for the heat kernel
 \begin{equation} \label{kernel}
 K( 0; t) = \frac{1}{2\pi L^2} \int_0^\infty d\lambda \tanh ( \pi \lambda) 
 e^{ -  t   (  \frac{\lambda^2}{L^2}  +   \frac{x^2}{L^2} ) }\, .
 \end{equation}
 The  partition function  is then given by 
 \begin{equation}
 \log Z =  \frac{1}{2} \int_0^\infty \frac{dt}{t}  \int dr d\theta \sqrt{g} K(0; t) \, .
 \end{equation}
 Here the integral over $AdS_2$ arises due to the fact we need to take the trace of the heat kernel to 
 evaluate one loop determinants. 
 Substituting the Kernel in (\ref{kernel})  and performing the $t$ integral we obtain 
 \begin{equation}
\log Z  =  - \frac{{\rm Vol} ( AdS_2) }{4\pi L^2} \int_0^\infty d\lambda \tanh ( \pi \lambda) 
 \log \left(\lambda^2  +   x^2  \right) \, ,
 \end{equation}
 where we  have ignored a $x$ independent constant resulting from $\log L^2$ in the integrand. 
 The integral over $\lambda$ needs to be UV regulated which we will do subsequently. 
 We substitute the regulated volume of $AdS_2$ which is given by \cite{Casini:2011kv,Klebanov:2011uf}
 \begin{equation}\label{volads}
 {\rm Vol} ( AdS_2)  = - 2\pi L^2 \, .
 \end{equation}
 Substituting this volume,  we obtain the 
 following result for the partition function 
 \begin{equation}
 \log Z(x)   =  \frac{1}{2} \int_0^\infty d\lambda \lambda  \tanh ( \pi \lambda) \log( \lambda^2+  x^2 ) \, .
 \end{equation}
 We can regulate this partition function by  first subtracting a $x$ independent constant, which is essentially the free energy
 in flat space. 
 We obtain the following free energy
 \begin{equation} \label{freads2}
 F(x) =  -\frac{1}{2}  \int_0^\infty d\lambda \Big[  \lambda \tanh(\pi  \lambda )\log( \lambda^2 +  x^2 )     - 
 \lambda \log ( \lambda^2)  \Big] \, .
 \end{equation}
This integral is now logarithmically divergent in the UV. 
We can evaluate this integral numerical by placing a cutoff at large  $\lambda$. 
Examining the integrand, we see that if sufficient derivatives in $x$ are taken, 
  there is a possibility of a divergence
 at $x=0$.  This divergence arises at the IR, that is at $\lambda \sim 0$ of the integrand. 
 This is manifest if  we  take the derivatives inside the integral. One sees that, 
 at  sufficient high order in derivatives with respect to  $x$ at $x=0$, the integrand diverges at $\lambda =0$. 
 However, to answer the question of whether the 
 sufficient derivatives of the free energy with respect to $x$ diverges at  $x=0$, we would need to do 
 the integral first and then differentiate with respect to $x$.  The  reason is that we do not wish to make an apriori assumption that
 the order of differentiation and integration can be interchanged, after all we are interested in the delicate issue of the 
 singular behaviour of the partition function under differentiation 
 \footnote{Though we were cautious and kept the order of differentiation to be after that of the integration, we have verified that the discontinuity 
 in the expectation value $\langle \phi^2\rangle$ is independent of the order of performing the differentiation.}.
 
\begin{figure}[h]
  \centering
  \begin{subfigure}[b]{0.4\linewidth}
    \includegraphics[height=.8\textwidth, width=1\textwidth]{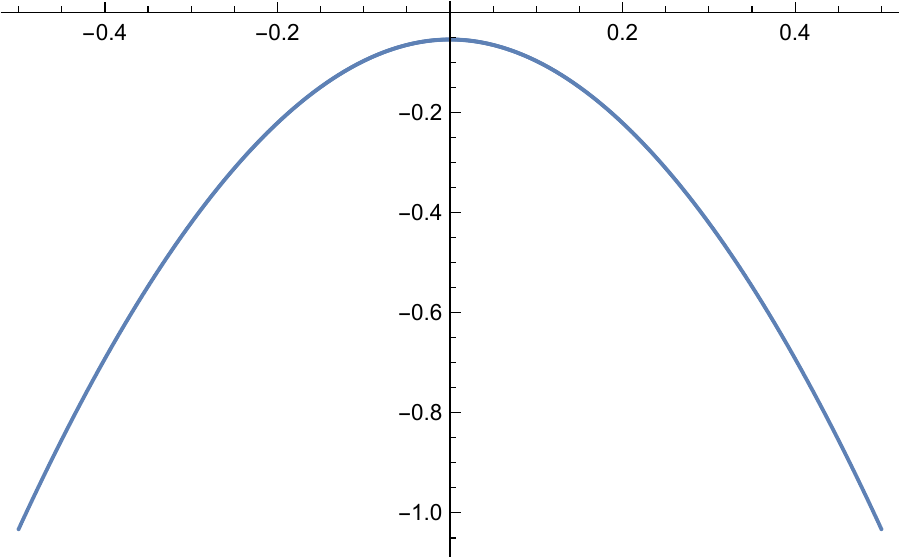}
    \caption{} \label{vardel1}
  \end{subfigure}
  \hspace{2cm}
  \begin{subfigure}[b]{0.4\linewidth}
    \includegraphics[height=.8\textwidth, width=1\textwidth]{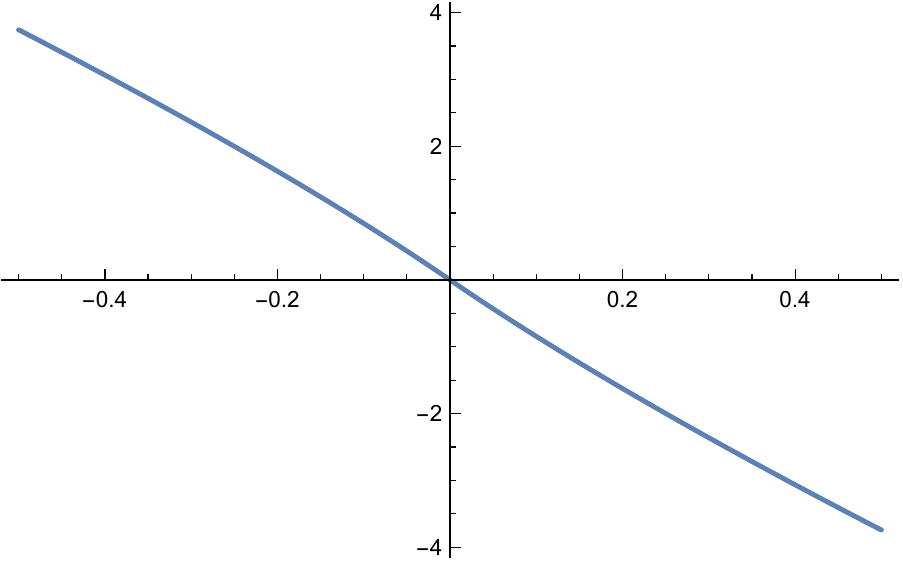}
    \caption{} \label{vardel2}
  \end{subfigure}
  
  \vspace{2cm}
   \begin{subfigure}[b]{0.4\linewidth}
    \includegraphics[height=.8\textwidth, width=1\textwidth]{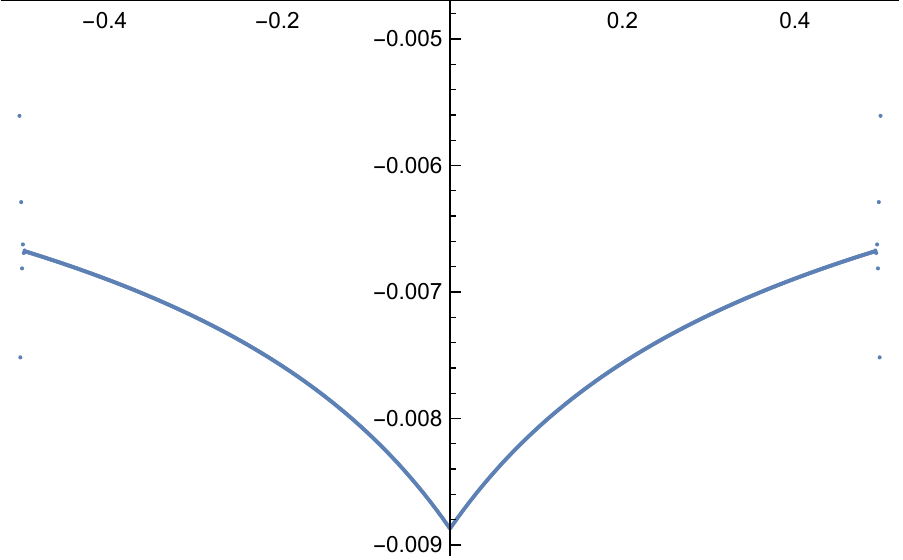}
    \caption{} \label{vardel3}
  \end{subfigure}
  \hspace{2cm}
  \begin{subfigure}[b]{0.4\linewidth}
    \includegraphics[height=.8\textwidth, width=1\textwidth]{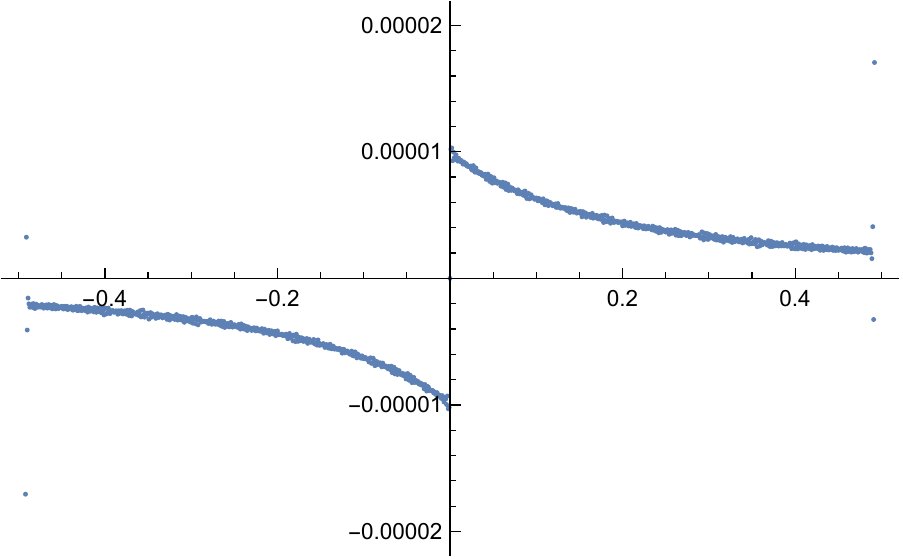}
    \caption{} \label{vardel4}
  \end{subfigure}
  \caption{  Plot of the  Free energy given of bosons on $AdS_2$ given  in (\ref{freads2})  and its derivatives. 
  (a) Plot of the regularized free energy $F(x)$ with respect to $x$. The integral is performed with a UV cut off at $\lambda =1000$. 
  The results do not change on further increasing this cut off. 
  (b) Plot of $F'(x) $, (c) Plot of $F''(x)$ finally (d) Plot of $F'''(x)$ }
  \label{fig1}
\end{figure}

 This question of whether the free energy is smooth 
 can be addressed numerically.  We have used Mathematica to perform the integral over $\lambda$ numerically
  with $x$ ranging from $-0.5$ to $0.5$ in steps of $0.001$ with a cut off on $\lambda = 1000$. 
 We then numerically differentiated, the resulting function.  
 The result  is shown in the figure \ref{fig1}.
 Note that as anticipated the resulting function in not smooth in $x$, 
 there is a kink  at $x=0$ at the 2nd derivative, and the 3rd derivative 
 is discontinuous at this point.  The results do not change on further increasing the cut off placed on $\lambda$.

From this simple analysis we observe that the partition  for bosons
 on $AdS_2$ with mass given in (\ref{frebmas}) is not a smooth function of $x$. 
 At  $x=0$,  when the mass saturates the 
 Breitenlohner-Freedman bound on $AdS_2$, the 3rd derivative of the function is discontinuous. 
 
\begin{figure}[!t]
	\centering
	\includegraphics[width=.6\textwidth]{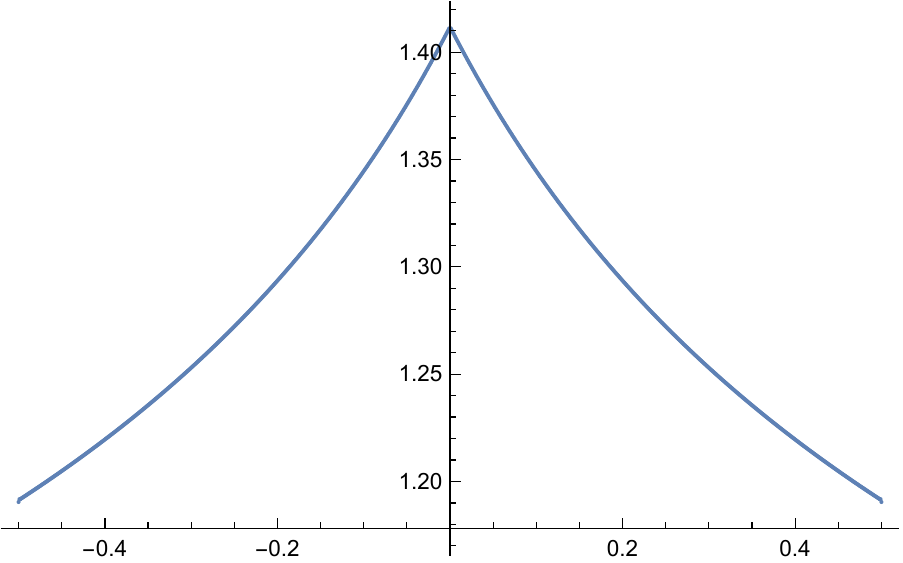}
	\caption{The  expectation value of $\phi^2$ given by (\ref{intexp1}) as a function of $x$  obtained numerically by differentiating the free energy.}
	\label{fig2}
\end{figure}

 It is also instructive to study the  derivative of the  partition function  with respective to $ m^2$, which 
 is related to the integrated value of the expectation value of $\phi^2$
 \begin{equation}\label{intexp}
-\frac{1}{2} \left\langle  \int dr d\theta \sqrt{g}  \phi^2( r, \theta ) \right\rangle = - \frac{d}{d m^2}  F (x)  =
-  \frac{L^2}{ 2 x}\frac{d}{dx} F(x) \, .
 \end{equation}
 Here we have used the action given in (\ref{adsaction}) and the mass in (\ref{frebmas}). 
 Since $AdS_2$ is a homogenous space, the expectation value is independent of the position in $AdS_2$. 
 Using this and the volume of $AdS_2$ given in (\ref{volads}) we obtain 
 \begin{equation} \label{intexp1}
 \langle \phi^2 \rangle|_{\rm numerics}  = -\frac{1}{2 \pi x}  \frac{d}{dx}  F (x) \, .
 \end{equation}
 The result of 
 evaluating this numerically is given in  figure \ref{fig2} . This expectation value clearly shows the 
 kink at  $x=0$  for which the mass saturates the BF bound. 

\subsection{The kink in $\langle \phi^2\rangle$  from $AdS_2$ Green's function}
\label{sec-greens}

The bilinear field $\phi^2$   acquires finite expectation values on $AdS_2$. 
In section \ref{sec1.1}  we have used the 
the  partition function of the scalar on $AdS_2$ differentiated with respect to  the mass squared, $ m^2$ and 
used the homogenity of $AdS_2$ to obtain the expectation value. 
We have seen that the slope of the  expectation value  $\langle \phi^2\rangle$ changes its sign at
the BF bound. 
 In this section we evaluate the change of slope analytically. 
The propagator on $AdS_2$ is known exactly. 
We can use the propagator and take the coincident limit to extract the expectation value of 
$\phi^2$ on $AdS_2$. 
We will see that though the expectation value is continuous in the variable $x$ which determines the mass
 (\ref{frebmas}), 
 the first derivative of the expectation value with respect to $x$  has  a discontinuity when the BF bound is saturated. 
 We will evaluate the change of this slope  analytically. 

To write down the propagator on $AdS_2$, it is useful to obtain the conformal dimension corresponding to the 
mass in (\ref{frebmas}).
This is given by  
\begin{eqnarray}\label{defdelta}
\Delta &=&  \frac{1}{2} + \sqrt{ ( m L)^2 + \frac{1}{4} } , \\ \nonumber
&=& \frac{1}{2} +  |x| \, .
\end{eqnarray}
Let us also define
\begin{eqnarray}
\nu &=& \sqrt{ ( m L) ^2 + \frac{1}{4} } , \\ \nonumber
&=& |x| \, .
\end{eqnarray}
Note that $\nu$ vanishes when the mass of the scalar saturates the BF bound. 
Let us define the distance
\begin{eqnarray}\label{defxi}
\frac{1}{\xi}=  \cosh r \cosh r' - \sinh r \sinh r' \cos( \theta - \theta') \, .
\end{eqnarray}
Then the scalar Greens function of this scalar is given by  \cite{DHoker:2002nbb}
\footnote{ See equation (6.12) of \cite{DHoker:2002nbb}}
\begin{eqnarray}
\langle \phi( r, \theta) \phi(r', \theta') \rangle &=&
G( \xi) = G(r,\theta, r',\theta')\,,  \\ \nonumber
 &=&
  \frac{ C_\Delta }{2\nu} \left( \frac{\xi}{2 } \right)^\Delta F( \frac{\Delta}{2}, \frac{\Delta}{2} + \frac{1}{2}, \nu + 1; \xi^2) \,
  \\ \nonumber
C_\Delta  &=& \frac{\Gamma( \Delta) }{ \sqrt{\pi } \Gamma ( \nu ) } \,.
\end{eqnarray}
Here $1/\xi = \cosh ( \mu/L)$, where $\mu$ is the geodesic distance between two points on $AdS_2$. 
The Greens function satisfies, the  equation 
\begin{equation}\label{diffeqg}
- \frac{1}{\sqrt{g} } \partial_\mu \Big( \sqrt{g} g^{\mu\nu} \partial_\nu G ( \xi) \Big)   +  m^2 G(\xi)   = 
\frac{1}{\sqrt{g}} \delta(r-r') \delta (\theta-\theta')\,.
\end{equation}
Note that from this equation and the metric in (\ref{metads2}), 
we see that the Greens function depends on the dimensionless variables, 
$\xi$ and the combination $ m^2 L^2$.

To obtain the expectation value  $\langle \phi^2 \rangle $ we take the coincident limit and then 
subtract the singular term. 
From the expression in (\ref{defxi}), we see that the  
 coincident limit is obtained by taking $\xi \rightarrow 1$, this leads to 
\begin{eqnarray}\label{adsgreen}
\lim_{\xi \rightarrow 1} G(\xi) &=& -  \frac{1}{4\pi}  \left[ 2 \gamma + \log( 2) + \log ( 1- \xi)  + \psi\big( \frac{1}{4} + \frac{|x|}{2} \big)  + 
\psi\big(\frac{3}{4} + \frac{|x|}{2}  \big)  \right]  \\ \nonumber
 && \qquad\qquad + O\Big[ ( 1- \xi) \log ( 1- \xi) , \;  (1-\xi) \Big] , \\ \nonumber
  &=&   -  \frac{1}{4\pi}  \Big[ 2 \gamma - \log 2 +  \log ( 1- \xi)  +  2 \psi\big( \frac{1}{2} + |x|\big) \Big]  +\cdots\,.
 \nonumber
\end{eqnarray}
In the first line of the above equation we have substituted for $\Delta$ using (\ref{defdelta}). To obtain the last line we have 
used the identity
\begin{equation}
\psi\big( \frac{1}{4} + \frac{|x|}{2} \big)  + 
\psi\big(\frac{3}{4} + \frac{|x|}{2}  \big)  = 2 \psi\big( \frac{1}{2} + |x|\big) - \log 4\,.
\end{equation}
We can remove the  term   $-\frac{1}{4\pi } \log ( 1 - \xi)$ in (\ref{adsgreen}),
 as it is singular and identify  the rest of the terms to be the 
expectation value $\langle \phi^2 \rangle $. 
However this definition is subject to an shift by an arbitrary constant. 
To relate to the next section, we take the $\xi \rightarrow 1$ limit, by the following limit. 
We set  $ r=r'$ and then take $r\rightarrow 0$. 
Then 
\begin{equation}
\lim_{r\rightarrow 0} \log( 1-\xi) |_{r=r'} = \log\Big[2r^2 \sin^2\big( \frac{\theta- \theta'}{2} \big) \Big] \,.
\end{equation}
Substituting this limit in (\ref{adsgreen}) , we obtain
\begin{equation}\label{greencompare}
\lim_{r\rightarrow 0} G(r, \theta, r, \theta') = -\frac{1}{2\pi}  \left( \gamma + \log\Big[r \sin\big( \frac{\theta- \theta'}{2} \big)\Big]   
+ \psi\big( \frac{1}{2} + |x|\big) \right) \,.
\end{equation}
It is important to note that the argument of the logarithm is dimensionless since $r$ is a dimensionless coordinate.
It is clear from the metric, that if we were to re-instate the dimensions of $r$, we would need to introduce the 
radius of $AdS_2$ in the logarithm. 
Let us identify the expectation value $\langle \phi ^2 \rangle$ as
\begin{equation} \label{expectation}
\langle \phi ^2 \rangle = -\frac{1}{2\pi} \Big[ \;\gamma +  \psi\big( \frac{1}{2} + |x|\big)\; \Big] \,.
\end{equation}
Though this identification is subject to an ambiguity by a constant, 
notice that the expectation value has a kink at  $x=0$,  its slope is discontinuous at this point. 
This discontinuity is unambiguous. 
We have
\begin{equation}\label{slopescalar}
\left. \frac{ d}{dx} \langle \phi ^2 \rangle \right|_{x>0} -  \left.  \frac{ d}{dx} \langle \phi ^2 \rangle \right|_{x<0}  = 
 -\frac{1}{2\pi} (\pi^2) = -\frac{\pi}{2} \,.
\end{equation}
Note that we have determined this change in slope in units of inverse radius of $AdS_2$.  If we define
\begin{equation}
\tilde m = \frac{x}{L}
\end{equation}
then we obtain
\begin{equation}\label{slopescalar1}
\left. \frac{ d}{d\tilde m } \langle \phi ^2 \rangle \right|_{\tilde m >0} - 
 \left.  \frac{ d}{d\tilde m } \langle \phi ^2 \rangle \right|_{\tilde m <0}  = 
 -\frac{1}{2\pi L} (\pi^2 ) = -\frac{\pi}{2 L} \,.
\end{equation}

\begin{figure}[!t]
	\centering
	\includegraphics[width=.6\textwidth]{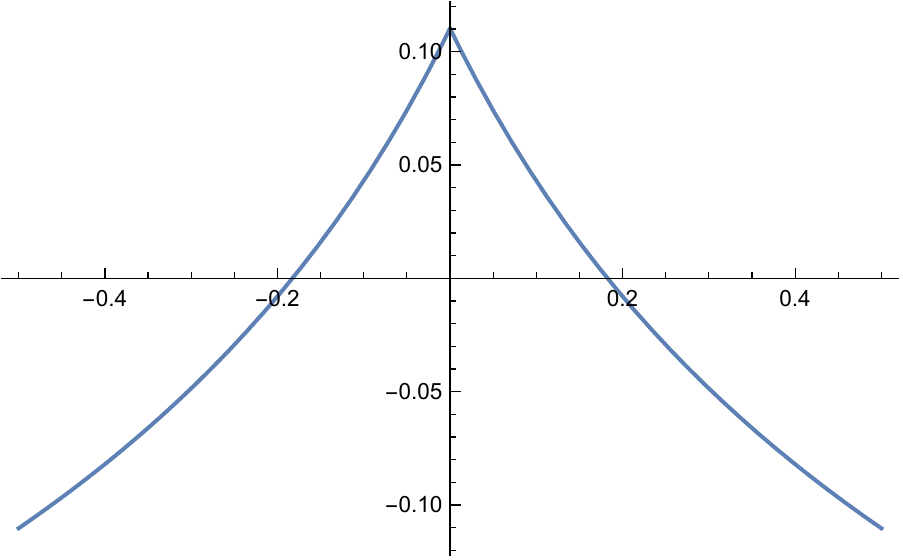}
	\caption{The expectation value of $\phi^2$ given by (\ref{expectation}) as a function of $x$.}
	\label{fig3}
\end{figure}

\begin{figure}[h]
  \centering
  \begin{subfigure}[b]{0.4\linewidth}
    \includegraphics[height=.8\textwidth, width=1\textwidth]{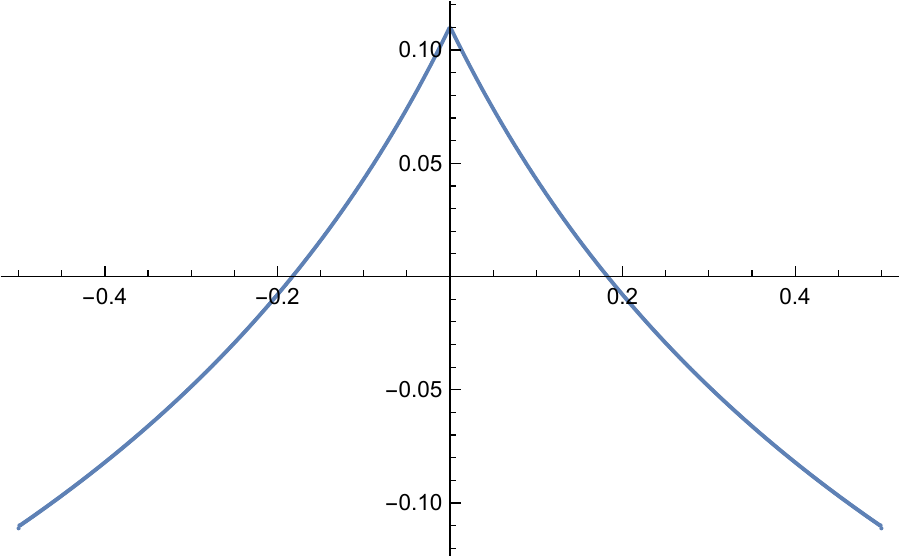}
    \caption{} \label{vardel11}
  \end{subfigure}
  \hspace{2cm}
  \begin{subfigure}[b]{0.4\linewidth}
    \includegraphics[height=.8\textwidth, width=1\textwidth]{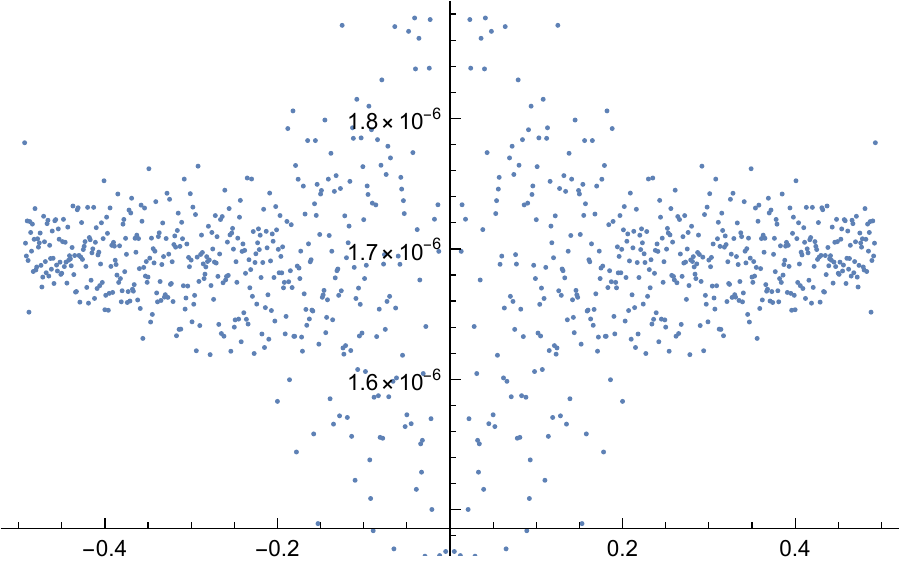}
    \caption{} \label{vardel22}
  \end{subfigure}
   \caption{ (a)  Plot of  $\langle \phi^2 \rangle_{\rm numerics}     + A  $ with 
   $A =-1.30159 $ with respect to $x$. 
   .  Note that the graph looks identically to  the one in
   Figure (\ref{fig3}). 
   (b) Plot of the difference  $\Delta$ as defined in \ref{difference}, observe that 
  $\Delta \leq 10^{-5}$. 
    }
  \label{fig4}
\end{figure}

Figure  \ref{fig3} contains the plot of the expectation value against $x$, note that this 
clearly shows a kink at $x=0$. 
One simple consistency check for the expectation value obtain in (\ref{expectation})  is the following:
using (\ref{intexp1}) we numerically evaluated the  expectation value of $\phi^2$, while in 
(\ref{expectation}) we have evaluated the expectation value analytically using the Greens function.
As we have emphasised that the definition of the expectation value is ambiguous upto a constant, therefore 
  we must have 
\begin{equation} \label{agree}
\langle \phi^2 \rangle_{\rm numerics}     + A =  \langle \phi^2 \rangle \,,
\end{equation}
where the left hand side of this equation is obtained by  (\ref{intexp1}) and $A$  is the constant which relates the 
two definitions. 
We have verified the equation (\ref{agree})  numerically
and found  $A = -1.30159$.  The LHS of (\ref{agree})   is plotted 
 in figure  (\ref{fig2}), it  precisely agrees with (\ref{fig3}). 
The  absolute value of the difference 
\begin{equation} \label{difference}
\Delta = \left|  \langle \phi^2 \rangle_{\rm numerics}     + A  - \langle \phi^2 \rangle \right| \,
\end{equation}
is plotted in 
figure (\ref{fig4}).
We see that $\Delta \leq 10^{-5}$. 

\subsubsection*{$\langle \phi^2 \rangle$ for the free boson on $\mathbb{R}^2$ }

At this point it is illustrative to compare the results for 
 the expectation value of the scalar bilinear  that we obtained on $AdS_2$ with that  
when one considers the free boson on flat space. 
The Green's function is given by 
\begin{eqnarray}
\langle \phi( \vec y )  \phi(0, 0) \rangle_{\mathbb{R}^2}  &=& 
\frac{1}{4\pi^2} \int  d^2 p \frac{  e^{i \vec p\cdot \vec y }}{ p^2 + m^2}\, , 
\\ \nonumber
&=& \frac{1}{2\pi} K_0( |m|  y ) \, .
\end{eqnarray}
Here $\vec y = (y_1, y_1)$ the co-ordinates on $\mathbb{R}^2$, $y^2 = \vec y\cdot \vec y$  and 
 $K_0$ refers to the modified Bessel function.
To obtain the expectation value we 
perform the short distance expansion on the Green's function we obtain 
\begin{eqnarray}
\lim_{\vec y \rightarrow 0} \langle \phi( \vec y ) , \phi(0, 0) \rangle_{\mathbb{R}^2}  = 
-\frac{1}{4\pi } \Big(2 \gamma - 2 \log 2  + \log ( m^2  y^2)  + O(m^2y^2, m^2y^2 \log( m^2y^2)  ) \Big) \, .
\nonumber \\
\end{eqnarray}
We can compare this  with equation (\ref{greencompare}), which is the result on $AdS_2$. 
There we see that the argument of the logarithmic divergence is dimensionless and the scale which is 
responsible for that is the radius of $AdS_2$.  On $\mathbb{R}_2$, the only scale is the mass. 
We can proceed by defining the expectation value by subtracting the logarithmic divergence just as it
was done in the $AdS_2$ case \footnote{We need to include the factor of $m^2$ in the logarithm since only 
then the argument is dimensionless for the theory on $\mathbb{R}_2$.},
we obtain 
\begin{equation}
\langle \phi^2 \rangle_{\mathbb{R}^2} = -\frac{1}{2\pi } ( \gamma -  \log 2 )\, . 
\end{equation}
This does not have a kink, more importantly it does not have a change in slope that we saw in the 
$AdS_2$ case in (\ref{slopescalar})  
\footnote{Even if the factor of $m^2$ was retained in the expectation value, the change 
in the slope of the expectation value is certainly not finite as in the case of $AdS_2$ seen in (\ref{slopescalar}).}.

\subsection{The kink in $\langle \phi^2 \rangle$ from its  Fourier  decomposition}
\label{sec-fourier}

In this section we  re-examine  two point function  $\langle \phi(r, \theta) \phi(r', \theta') \rangle$ in terms of 
its Fourier decomposition along the angular direction in $AdS_2$. 
We will see that the discontinuity in the slope of the expectation value $\langle \phi^2 \rangle$ arises solely  due to the 
zero mode in the angular direction. 
As before,  we  parametrise the mass of the scalar as 
\begin{equation} \label{masssq}
m^2 = -\frac{1}{4 L^2} + \frac{ x^2}{L^2}\,,
\end{equation}
therefore the Breitenlohner-Freedman bound is reached at $x=0$. 
Each Fourier mode of the  Green's function  $\langle \phi(r, \theta) \phi(r', \theta') \rangle$ satisfies 
an ordinary differential equation in the radial co-ordinate $r$ with a delta function source. 
The Green's function have to satisfy smoothness at the origin $r=0$ and should be normalizable at the boundary. 
We will see that though the mass depends on $x^2$, the choice of the normalizable wave function to construct the 
Green's function depends on the sign of $x$. 
The kink in  the expectation value at $x=0$ results due to the different choice of wave functions for
$x>0$ versus $x<0$ for the zero mode along the angular direction.

To begin,  the Green's function satisfies the differential equation (\ref{diffeqg})
\begin{eqnarray}\label{greeneq}
&& \partial_r \big[ \sinh r\, \partial_r G(r, \theta, r', \theta') \big]  +\frac{1}{\sinh r} \partial_\theta^2 G(r, \theta, r', \theta') 
   \nonumber \\
&& \qquad\qquad\qquad + \frac{1}{4} ( 1- 4 x^2)  \sinh r\;  G(r, \theta, r', \theta') =   -\delta( r- r') \delta(\theta - \theta')  \,.
\end{eqnarray}
We Fourier expand the Green's function as 
\begin{equation}
G(r, \theta, r', \theta') =\frac{1}{2\pi} \sum_{p=-\infty}^\infty \hat G_{p} ( r, r', p ) e^{i p ( \theta - \theta') } \,.
\end{equation}
Then each Fourier mode of the Green's function satisfies the equation
\begin{eqnarray}
&& \sinh r\, \partial_r^2 \hat G(r, r' , p ) + \cosh r\, \partial_r \hat G (r, r', p) - \frac{p^2}{\sinh r}  \hat G  (r, r', p)   \\ \nonumber
&&  \qquad\qquad\qquad\qquad\qquad +\frac{1}{4}  ( 1- 4x^2)  \sinh r \, \hat G (r, r', p )  =- \delta ( r -r')  \,.
\end{eqnarray}
To construct the Green's function, we solve the homogenous equation. Let 
$\varphi_{\rm in }( r, p)$ be the solution which is smooth at the origin and let $\varphi_{\rm out}(r, p)$  be the solution 
which obeys normalizable boundary conditions as $r\rightarrow \infty$ \footnote{
These solutions behave as $e^{-(\frac{1}{2} + \delta)r}$ with $\delta >0$ as $r\rightarrow \infty$. Such a behaviour
ensures that the integral $\int dr d\theta \sqrt{ {\rm det} g} \varphi^2 $ is well behaved at radial infinity. }
Then the Green's function is given by 
\begin{equation} \label{greenfdef}
\hat G(r, r', p) =
\begin{cases}
 \frac{1}{W_p}  \varphi_{\rm in} (r) \varphi_{\rm out} ( r')  &  {\rm for} \; r<r', \\
 \frac{1}{W_p} \varphi_{\rm out} ( r) \varphi_{\rm in} (r')  &{\rm for } \; r>r' \,.
 \end{cases}
\end{equation}
Here $W_p$ is a constant related to the Wronskian of the homogenous differential equation. 
By construction, the Green's function is continuous at $r=r'$. 
The discontinuity of the slope at $r=r'$ is obtained by integrating the equation (\ref{greeneq}) in this neighbourhood. 
This results in the equation
\begin{equation}\label{defwp}
\frac{\sinh r }{W_p}\left(  \frac{d\varphi_{\rm out } (r, p ) }{dr} \varphi_{\rm in} (r, p ) -  
\frac{d\varphi_{\rm in } (r, p ) }{dr} \varphi_{\rm out} (r, p ) 
\right) =- 1 \,.
\end{equation}

Since we are interested in the expectation value $\langle \phi^2 \rangle$ which is obtained in the 
coincident limit, we can examine the solutions $\varphi_{\rm in }(r, p), \varphi_{\rm out}(r, p)$
as series expansions as $r\rightarrow 0$ and construct the Green's function in the coincident limit.

\subsubsection*{Non-zero modes}

We will first construct the solutions for $p\neq 0$ as a series expansion in $r$. 
The smooth solutions at $r=0$ are given by 
\begin{eqnarray} \label{smoothsol}
\varphi_{\rm in}  (r, p ) =  r^{p}\Big(1-\frac{1}{48(1+p)}(3+4p+4p^{2}-12x^{2})r^{2} + \cdots \Big),\qquad \text{for}\quad p\geq 1 , \\ \nonumber
\varphi_{\rm in } (r, p ) = r^{-p}\Big(1-\frac{1}{48(1-p)}(3-4p+4p^{2}-12x^{2})r^{2}+ \cdots \Big),\qquad \text{for}\quad p =1 \,.
\end{eqnarray}
To obtain $W_p$ which is determined by (\ref{defwp})
we need the solution which is normalizable at $r\rightarrow \infty$. 
Using uniqueness of the solution of the differential equation, we can write this solution as 
 a linear combination 
of the singular solution at $r\rightarrow 0$ and an independent solution, which is smooth at $r\rightarrow 0$. 
From the Frobenius series expansion, these solutions are given by
\begin{eqnarray} \label{singsol}
&&\varphi_{\rm out} ( r, p) =r^{-p}\Big(1-\frac{(3-4p+4p^{2}-12x^{2})}{48(1-p)}r^{2} +\cdots \Big) \\ \nonumber
&& \qquad\qquad\qquad + r^p \, C_{( p) }\ln r\,\Big(1-\frac{(3+4p+4p^{2}-12x^{2})}{48(1+p)}r^{2} +\cdots \Big),\,\, \quad \text{for}\quad p> 1\,,\nn\\
&&\varphi_{\rm out} (r, 1)=\frac{1}{r}\Big(1+\frac{1+1560x^{2}-2160x^{4})}{480(11-12x^{2})}r^{2}+ \cdots \Big)  \nonumber\\ \nonumber
&& \qquad\qquad\qquad  +
r \Big(\frac{4x^{2}-1}{8}\ln r+
C_{(1) }\Big)\Big(1-\frac{(11-12x^{2})}{96}r^{2} +\cdots \Big),\nn\\
&&\varphi_{\rm out}(r, p) 
=r^{p}\Big(1-\frac{(3+4p+4p^{2}-12x^{2})}{48(1+p)}r^{2} +\cdots \Big) \nonumber \\ 
&&\qquad\qquad\qquad    +
r^{-p} \, C_{(p) }\ln r \,\Big(1-\frac{(3-4p+4p^{2}-12x^{2})}{48(1-p)}r^{2} +\cdots \Big),\,\, \quad \text{for}\quad p< -1\,,\nn\\
&&\varphi_{\rm out}(r, -1)
=\frac{1}{r}\Big(1+\frac{1+1560x^{2}-2160x^{4})}{480(11-12x^{2})}r^{2}+\cdots \Big) \nonumber \\
&& \qquad\qquad\qquad  +r\, \Big(\frac{4x^{2}-1}{8}\ln r+ C_{(-1) } \Big)\,\Big(1-\frac{(11-12x^{2})}{96}r^{2}+\cdots 
\Big)\,. \nonumber
\end{eqnarray}
Here $C_{(p)}$ refer to  undermined constants which can be fixed by extrapolating the normalizable solution 
at $r\rightarrow \infty$ to the origin. 
We will see that we would not need the constants $C_{(p)}$, to determine the behaviour of the Greens function 
as $r\rightarrow 0$. 
The solution to the  Wronskian of the differential equation is given by 
\begin{equation}
\sinh r \left(  \frac{d\varphi_{\rm out } (r, p ) }{dr} \varphi_{\rm in} (r) -  \frac{d\varphi_{\rm in } (r, p ) }{dr} \varphi_{\rm out} (r, p ) 
\right)  = - W_p \,,
\end{equation}
where $W_p$ is a constant. 
This constant can be fixed by examining the Wronskian at $r\rightarrow 0$. 
Now plugging in the solutions in (\ref{smoothsol}) and (\ref{singsol}) for $|p| \geq 1$, we obtain 
\begin{equation}\label{valwp}
W_p = 2|p| , \qquad  {\rm for}\; |p| \geq 1 \,,
\end{equation}

\subsubsection*{Zero mode}

Let us first construct the smooth solution near $r=0$ using the Frobenius expansion
This is given by 
\begin{equation}
\varphi_{\rm in} (r, 0)   = 1- \frac{1-4x^2}{16}  r^2 + \cdots \,.
\end{equation}
Again using the Frobenius expansion, we can write the solution which is singular  at the origin, but normalizable 
at $r\rightarrow\infty$  as
\begin{equation} \label{sinzero}
\varphi_{\rm out} (r, 0) =  (C_{(0)}+\ln r)(1-\frac{3-12x^{2}}{48}r^{2})-(\frac{1}{48}+\frac{x^{2}}{4})r^{2} +\cdots \,.
\end{equation}
For the zero mode we would need the constant $C_{(0)}$. 
As we have mentioned earlier, this constant can be fixed by obtaining the solution which is normalizable 
at $r\rightarrow \infty$ and extrapolating it to the origin. 
The homogenous differential equation can be solved exactly for $p=0$. 
The  two independent solutions are given by 
\begin{equation}\label{solnp0}
y^{\frac{1}{4} (1-2x) }\,{}_{2}F_{1}(\frac{1}{4}-\frac{x}{2} ,\frac{3}{4}-\frac{x}{2};1-x;y), 
\qquad 
y^{\frac{1}{4} (1+2x) }\, {}_{2}F_{1}(\frac{1}{4}+\frac{x}{2} ,\frac{3}{4}+\frac{x}{2};1+x;y) \,,
\end{equation}
where 
\begin{equation}
y =\frac{1}{(\cosh r)^2 } \,.
\end{equation}
It is clear that the normalizable solution  at $r\rightarrow \infty $ depends on the sign of $x$. 
For $x>0$ the solution is given by 
\begin{equation}\label{zeromodesol}
\left. \varphi_{\rm out} (r, 0) \right|_{x>0}  = - \frac{ \Gamma( \frac{1}{4} + \frac{x}{2} ) \Gamma( \frac{3}{4} + \frac{x}{2} ) }{ 2 \Gamma( 1+x) } 
y^{\frac{1}{4} (1+2x) }{}_{2}F_{1}(\frac{1}{4}+\frac{x}{2} ,\frac{3}{4}+\frac{x}{2};1+x;y) \,.
\end{equation}
Here we see that the solution behaves as 
\begin{equation}
\left. \lim_{r\rightarrow\infty} \varphi_{\rm out} (r, 0)\,\right|_{x>0}  \sim e^{ -( \frac{1}{2} + x)r}  \,.
\end{equation}
This is the required behaviour in $AdS_2$ for the solution to be normalizable. 
Expanding this solution at the origin we can determine the constant $C_{(0)}$ which is given by 
\begin{equation}\label{c01}
C_{(0)}|_{x>0} = \psi\big( \frac{1}{2} + x\big) + \gamma - \log 2 \,,
\end{equation}
Note that the overall normalization for  the solution  in (\ref{zeromodesol}) is chosen so that the coefficient of $\log(r)$ in the 
$r\rightarrow 0$ limit is unity as in the Frobenius expansion (\ref{sinzero}). 
Similarly  we see for $x<0$, the normalizable solution  is given by 
\begin{equation}
\left. \varphi_{\rm out} (r, 0) \right|_{x<0}  =  - \frac{ \Gamma( \frac{1}{4} - \frac{x}{2} ) \Gamma( \frac{3}{4} - \frac{x}{2} ) }{ 2 \Gamma( 1-x) } 
y^{\frac{1}{4} (1-2x) }{}_{2}F_{1}(\frac{1}{4}-\frac{x}{2} ,\frac{3}{4}-\frac{x}{2};1-x;y), 
\end{equation}
and expanding this solution at the origin we find
\begin{equation}\label{c02}
C_{(0)}|_{x<0} = \psi\big( \frac{1}{2} - x \big) + \gamma - \log 2 \,.
\end{equation}
Evaluating the proportionality constant $W_0$ for the Wronskian of the solutions for $p=0$ by using their expansion 
at $r=0$ we find that 
\begin{equation} \label{valwp0}
W_0 = -1\,.
\end{equation}

\subsubsection*{Greens function}

Substituting the solutions in (\ref{smoothsol}), (\ref{singsol}), (\ref{solnp0}), (\ref{zeromodesol})
 and value of the constant $W_p$ (\ref{valwp}), (\ref{valwp0})  into the expression for each mode of the 
Green's function given
in (\ref{greenfdef}), we obtain the following expressions when $r=r'$ and in the small $r$ expansion
\begin{eqnarray}\label{greenfnmode}
\hat G( r, r, p) &=& \frac{1}{2p}-\frac{1-4x^{2}}{16p(1-p^{2})}r^{2}+ \cdots , \qquad p>1\nn\\
\hat G(r, r, 1) &=&  \frac{1}{2}-
\frac{ (151-720x^{2}+720x^{4}+(165-840x^{2}+720x^{4})\ln r-(1320-1440x^{2})C_{(1)} )}{240(11-12x^{2})}r^{2}\nn \\
& & \qquad\qquad+  O(r^{4})\,,\nn\\
\hat G(r, r, 0) &=& -\ln r-C_{0}+\frac{1}{48}(1+12x^{2}+6(1-4x^{2})\ln r+6(1-4x^{2})C_{(0)})r^{2} + \cdots \,,\nn\\
\hat G(r, r, -1) &=& \frac{1}{2}-\frac{151-720x^{2}+720x^{4}+(165-840x^{2}+720x^{4})\ln r-(1320 -1440x^{2}) C_{(-1)}}{240(11-12x^{2})}r^{2} \nn \\
&&\qquad\qquad + O(r^4)  \,,\nn\\
\hat G(r, r, p) &=&-\frac{1}{2p}+\frac{1-4x^{2}}{16p(1-p^{2})}r^{2}+\cdots\,, \qquad p<-1
\end{eqnarray}
It is clear from these expressions that the constants $C_{(p)}$  for $|p| \geq 1$ do not matter for the leading behaviour 
in the small $r$ expansion. 
Considering the leading terms for each $p$  (\ref{greenfnmode}), we can re-sum the Fourier coefficients
\begin{equation} \label{fmgreen}
\lim_{r\rightarrow 0} G( r,\theta, r', \theta') = - \frac{1}{2\pi} \left( \log( 2r \sin\frac{(\theta-\theta')}{2} )   + C_{(0)} \right)
+ \cdots \,.
\end{equation}
The constant $C_{(0)}$ depends on the sign of $x$ and is given in (\ref{c01}), (\ref{c02}). 
Comparing (\ref{greencompare}) and   (\ref{fmgreen}),  
we see that the constant precisely agrees and 
removing the singular term we obtain 
\begin{equation} \label{expectation2}
\langle \phi ^2 \rangle = -\frac{1}{2\pi} \Big[ \;\gamma +  \psi\big( \frac{1}{2} + |x|\big)\; \Big]  \,.
\end{equation}
The  insight we gained by doing the mode by mode analysis of the Green's function is that 
the discontinuity in the slope of the Green's function is entirely due to the $p=0$ mode and the reason it occurs 
is because the condition that the wave function is normalizable as $r \rightarrow \infty$ 
depends on the sign of $x$, which parametrises the difference of the mass squared from the 
BF bound (\ref{masssq}). 
Therefore every time the mass of the boson
 crosses the BF bound there is a kink in the expectation value of  $\langle \phi^2 \rangle$ 
 and its slope changes sign.

 \section{Fermions on $AdS_2$} \label{fermion}

 In this section we study the partition function of fermions on $AdS_2$ as we vary their mass. 
 Again we first do this numerically and then we  evaluate the expectation value of the fermion bilinear by obtaining the 
 Green's function and taking the coincident limit. 
 
 \subsection{The free energy and  fermion bilinear: numerics}
 
In this section, we repeat the previous analysis in the case of a free Dirac fermion on AdS$_{2}$. 
There are 2 actions we will consider, the first is that of a massive Dirac fermion 
\begin{equation} \label{act1}
S_I=\int d^{2}x\sqrt{g}\,\bar\psi ( \slashed D+ m )\psi\,.
\end{equation}
while the second action is given by 
\be \label{act2}
S_{II}=\int d^{2}x\sqrt{g}\,\bar\psi ( \slashed D+ m\g_{t})\psi\,.
\ee
The gamma matrices are given by 
\be
\g_{t}=\begin{pmatrix}1&0\\0&-1\end{pmatrix},\quad \g^{1}=\begin{pmatrix}0&1\\1&0\end{pmatrix},\quad \g^{2}=\begin{pmatrix}0&i\\-i&0\end{pmatrix} \,,
\ee
while the vielbein are 
\begin{equation}
e^{1}=L\,dr, \qquad\qquad\qquad  e^{2}=L\sinh r\,d\theta \,.
\end{equation}
The direction $1, 2$ correspond to directions $r, \theta$ respectively. 
The action $S_I$ is the canonical Dirac action with a mass.  $S_{II}$ is the action that occurs on 
Kaluza-Klein reduction of fermions on $AdS_2\times S^1$ or $AdS_2\times S^2$ as we will see in subsequent sections. 
We parametrize the mass as 
\begin{equation}
m = \frac{x}{L}
\end{equation}
The BF bound for fermions  is  given by the 
condition  \cite{Amsel:2008iz,Dias:2019fmz} \footnote{
The BF bound for fermions is independent of dimensions, see equation B.19 of  \cite{Amsel:2008iz} , equation 4.3 of 
\cite{Dias:2019fmz}}. 
\begin{equation} \label{BFfermion}
m^2 \geq 0 \,.
\end{equation}
We will see for  both these actions (\ref{act1}), (\ref{act2}), the partition function is not smooth at $x=0$, that is when the 
mass saturates the BF bound. 
We will also show that for  $S_I$, the expectation value  the fermion bilinear $\bar \psi \psi$ is discontinuous  at $x=0$, while 
for $S_{II}$, the  expectation value of $\bar\psi \gamma_t\psi$ is discontinuous.

To evaluate the partition function it is useful to introduce the normalizable  eigen functions of the Dirac operator on $AdS_2$. 
These wave functions were constructed in \cite{Camporesi:1994ga}, they are given by 
\bea \label{spinor1}
\chi^{\pm}_{p}(\lambda) &=&\frac{1}{\sqrt{4\pi L^{2}}}\Big|\frac{\Gamma(1+p+i\lambda)}{\Gamma(p+1)\Gamma(\frac{1}{2}+i\lambda)}\Big|\,e^{i(p+\frac{1}{2})\theta} \\ \nn
&&\begin{pmatrix}i\frac{\lambda}{p+1}\cosh^{p}\frac{r}{2}\sinh^{p+1}\frac{r}{2}\;{}_{2}F_{1}(p+1+i\lambda,p+1-i\lambda;p+2;-\sinh^{2}\frac{r}{2})\\\pm\cosh^{p+1}\frac{r}{2}\sinh^{p}\frac{r}{2}\;{}_{2}F_{1}(p+1+i\lambda,p+1-i\lambda;p+1;-\sinh^{2}\frac{r}{2})\,\end{pmatrix} \,.
\eea
Here 
\begin{equation} \label{domain}
p \in \mathbb{Z}, \qquad   p\geq 0, \qquad 0 <\lambda <\infty
\end{equation}
These eigen functions satisfy 
\begin{equation}
\slashed D\chi_{\pm,\,p}(\lambda)=\pm \frac{i\lambda}{L}\chi_{\pm,\, p}(\lambda) \,.
\end{equation}
Similarly,  to cover negative half integer quantum number along $\theta$ eigen states, we have
\bea \label{spinor2}
\eta^{\pm}_{p}(\lambda)&=&\frac{1}{\sqrt{4\pi L^{2}}}\Big|\frac{\Gamma(1+p+i\lambda)}{\Gamma(p+1)\Gamma(\frac{1}{2}+i\lambda)}\Big|\,e^{-i(p+\frac{1}{2})\theta}\\ \nn
&&\begin{pmatrix}\cosh^{p+1}\frac{r}{2}\sinh^{p}\frac{r}{2}\;{}_{2}F_{1}(p+1+i\lambda,p+1-i\lambda;p+1;-\sinh^{2}\frac{r}{2})\\\pm i\frac{\lambda}{p+1}\cosh^{p}\frac{r}{2}\sinh^{p+1}\frac{r}{2}\;{}_{2}F_{1}(p+1+i\lambda,p+1-i\lambda;p+2;-\sinh^{2}\frac{r}{2})\,\end{pmatrix} \,.
\eea
Here too $p, \lambda$ takes values as given in (\ref{domain}). 
These eigen functions satisfy
\begin{equation}
\slashed D\eta_{\pm,\, p}(\lambda)=\pm \frac{i\lambda}{L}\eta_{\pm, \, p }(\lambda) \,.
\end{equation}
Note that there exists distinct eigen functions for both positive and negative $\lambda$. 

Let us examine the action $S_{I}$.  Performing the path integral  we obtain 
\begin{equation}
Z_I =   {\rm Det} ( \slashed D +   m) \,.
\end{equation}
Now since the eigen values  of $  \slashed D$ range over both positive and negative $\lambda$, we can see that 
\begin{equation}
{\rm Det} (\slashed D + m)  =  {\rm Det} (  \slashed D -  m)  \,.
\end{equation}
Therefore we can write the partition function as 
\begin{equation} \label{partspin1}
Z_I = \sqrt{ {\rm Det} ( -\slashed D^2 +m^2  ) } \,.
\end{equation}
Therefore we consider the heat kernel 
\begin{equation}
K_I ( r, \theta, r', \theta' ; t) = \sum_{p=0}^\infty \int_0^\infty d\lambda e^{ - t( \frac{\lambda ^2 }{L^2} + \frac{x^2}{L^2} )  } 
\sum_{a = \pm} \left( \bar \chi_{p}^{a } ( \lambda)  \chi_{p}^{a } ( \lambda)  + 
\bar \eta_{p}^{a } ( \lambda)  \eta_{p}^{ a} ( \lambda) \right)  \,.
\end{equation}
To evaluate the one loop determinant, we need to take the coincident limit of the heat kernel. Again, as $AdS_2$ is a homogenous
space, we can take the coincident limit at the origin. 
The wave functions in (\ref{spinor1}), ( \ref{spinor2}) have he property that only $p =0$ is non-vanishing at $r=0$ and 
\begin{equation}
\sum_{a = \pm} \left( \bar \chi_{0}^{a } ( \lambda)  \chi_{0}^{ a} ( \lambda)  + 
\bar \eta_{0}^{a } ( \lambda)  \eta_{0}^{ a} ( \lambda) \right) =
\frac{1}{\pi L^2}   \lambda \coth (\pi \lambda)  \,.
\end{equation}
Substituting this value of the wave functions in  the coincident limit of the heat kernel, we obtain 
\begin{equation} \label{coinci}
K_I( 0; t) =  \frac{1}{\pi L^2} \int_0^\infty d\lambda  \coth (\pi \lambda) e^{ - t( \frac{\lambda ^2 }{L^2} + \frac{x^2}{L^2} )  } \,.
\end{equation}
Using this expression for the coincident heat kernel, the expression for the partition function $Z_I$ is given by 
\begin{equation}
\log Z_I =   -\frac{1}{2 \pi L^2} \int_0^\infty \frac{dt}{t} \int dr d\theta \sqrt{g} K(0;t)  \,.
\end{equation}
Substituting the coincident limit  of the kernel from (\ref{coinci}), we obtain 
\begin{equation}
\log Z_I =  \frac{ {\rm Vol}( AdS_2) }{ 2\pi L^2}   \int_0^\infty d\lambda  \lambda \coth (\pi \lambda) 
\log( \lambda^2 + x^2) \,.
\end{equation}
Here we have ignored the $x$ independent constant, substituting the regularised volume of $AdS_2$ and again regularising 
by  removing the $x$ independent  free energy in flat space we obtain for the the free energy
\begin{equation}
F_I(x) = \int_0^\infty d\lambda \Big[ \lambda \coth(\pi \lambda) \log ( \lambda^2 + x^2)  - \lambda \log( \lambda^2) 
\Big] \,.
\end{equation}
We can obtain the expectation value of the bilinear $\bar \psi \psi$ by differentiating the free energy with respect to $x$
\begin{equation}
-\langle  \int dr d\theta \sqrt{g} \bar\psi \psi \rangle =  - L\frac{d}{dx} F_I(x)  \,.
\end{equation}
Since $AdS_2$ is a homogenous space,  we can obtain the expectation value of the fermion bilinear by factoring out the 
volume of $AdS_2$. 
\begin{equation}
\langle \bar \psi \psi \rangle = - \frac{1}{2\pi L} \frac{d}{dx} F_I(x)  \,.
\end{equation}

Now let us evaluate the partition function corresponding to the action $S_{II}$ in (\ref{act2})
\begin{equation}
Z_{II}  = {\rm det} (  \slashed D + m \gamma_t) \,.
\end{equation}
Here we write the partition function as 
\begin{equation} \label{partspin2}
Z_{II}   = \sqrt{ {\rm det} (  \slashed D + m \gamma_t)^2}  =\sqrt{  {\rm det} (  - \slashed D^2 + m ^2 ) } \,.
\end{equation}
Comparing (\ref{partspin1})  and (\ref{partspin2}), we see that  both the partition function are identical. 
Therefore we have
\begin{equation} \label{free2act}
F_{II}(x) = \int_0^\infty d\lambda \Big[ \lambda \coth(\pi \lambda) \log ( \lambda^2 + x^2)  - \lambda \log( \lambda^2) 
\Big] \,.
\end{equation}
Now from the action $S_{II}$ we see that we can obtain the expectation value of the bilinear $\bar\psi \gamma_t \psi$ by 
differentiating with respect to $x$. 
This leads to 
\begin{equation}
-\langle  \int dr d\theta \sqrt{g} \bar\psi \gamma _t \psi \rangle =  - L\frac{d}{dx} F_{II}(x)  \,.
\end{equation}
Again using the homogenity of  $AdS_2$, we obtain 
\begin{equation} \label{bilinearexp}
\langle \bar \psi \gamma_t \psi \rangle =  -\frac{1}{2\pi L} \frac{d}{dx} F_{II}(x) \,.
\end{equation}
Since the free energies of both the actions are identical, the expectation value of the bilinears $\bar\psi \psi$ for the theory with the 
action $S_I$ is the same as that of $\bar \psi \gamma_t \psi$ for the theory with action $S_{II}$. 

\begin{figure}[!t]
	\centering
	\includegraphics[width=.6\textwidth]{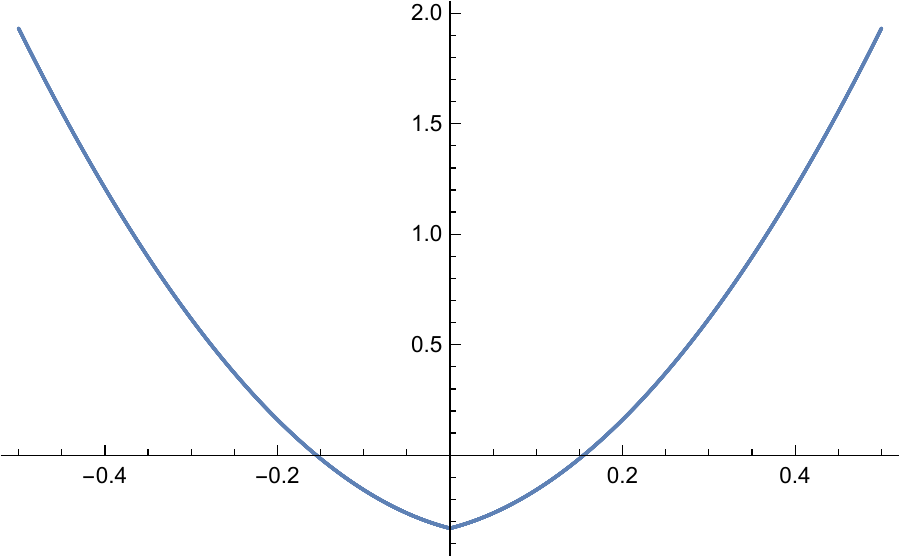}
	\caption{The free energy of fermions on $AdS_2$ given in (\ref{free2act})  as a function of the mass parametrized by $x$. 
	 }
	\label{fig5}
\end{figure}

\begin{figure}[!t]
	\centering
	\includegraphics[width=.6\textwidth]{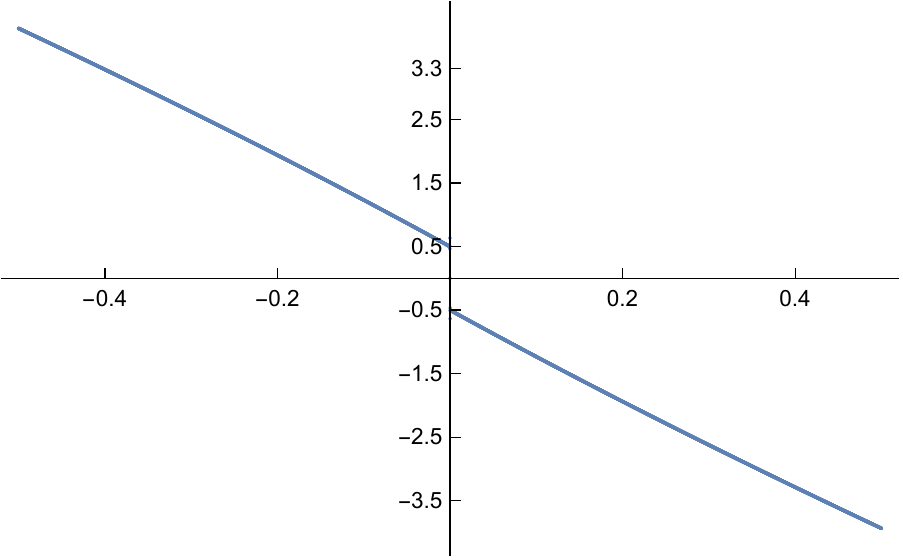}
	\caption{Expectation value  $\pi L \langle \bar \psi \gamma_t \psi\rangle$ as 
	 function of the mass parametrized by $x$.  Note that the expectation value is discontinuous at $x=0$, which is the 
	 BF bound for fermions.  From the graph we can read out that the jump in  
	 $\pi L \langle \bar \psi \gamma_t \psi\rangle$  at the BF bound is $-1$. }
	\label{fig6}
\end{figure}

Let us now proceed to evaluate the free energy numerically. 
Since the partition function for both the actions are same, for concreteness we will use the action $S_{II}$. 
We can evaluate the integral in (\ref{free2act}) numerically with a cutoff at $\lambda =1000$.  We run $x$ from 
$-0.5$ to $0.5$ in steps of $0.0001$. 
The result is shown in figure \ref{fig5}, note that the plot demonstrates a kink in the free energy at the BF bound $x=0$. 
Numerically differentiating the free energy using the formula in (\ref{bilinearexp}) we obtain the result shown in 
figure \ref{fig6} which indicates that there is a jump in the expectation value of the fermion bilinear 
$\langle \bar \psi \gamma_t \psi \rangle$
at the BF bound $m=0$.  Evaluating this jump numerically we find 
\begin{equation} \label{expectjump}
\pi L \left. \Big( \langle \bar \psi \gamma_t \psi\rangle|_{m\rightarrow 0^+ } - \langle \bar \psi \gamma_t \psi\rangle|_{m\rightarrow 0^- }
\Big)\right|_{\rm numerics}  = -1.005\,.
\end{equation}
This jump can be clearly seen in the figure \ref{fig6}. 

\subsection{Fermion bilinear from the Greens function}

We can compute the expectation value  $\langle \bar\psi\gamma_t\psi\rangle $ as well as $\langle \bar \psi \psi \rangle$  
using the Green's function 
of the fermion.  
Given the Greens's function we can use the point split approach to evaluate these expectation values just as in the 
case of bosons. 
Therefore, let us consider the Green's function corresponding to the action (\ref{act2}). 
The equation of motion for the Green's function  is given by 
\begin{equation} \label{eqgreenf}
\left[ ( i \slashed D  + m \gamma_t)  G(x, x') \right]_{\alpha \beta}  = \frac{\delta^{2}(x-x')}{\sqrt{g(x)}}\delta_{\alpha\beta} \,.
\end{equation}
Here $\alpha, \beta$ refer to spinor indices. 
To solve for the Green's function we adapt the method of \cite{Mueck:1999efk}. 
We make the following  ansatz
\begin{equation} \label{ansatz}
G(x, x') = \left[ \hat\alpha (\mu ) \gamma_t + \hat\beta(\mu) n_\nu \gamma^\nu ) \right] \Lambda( x, x') \,,
\end{equation}
where $\mu(x, x')$ is the geodesic distance between points $x$ and $x'$ given  by 
\begin{equation} \label{defmuxi}
\cosh\big( \frac{\mu}{L} \big) = \frac{1}{\xi}  \,,
\end{equation}
with $\xi$ given in (\ref{defxi}). 
$n_\nu$ is the unit norm  tangent vector at the end point $x$ of the geodesic, defined  by 
\begin{equation}
n_\nu = D_\nu  \mu ( x, x'), \qquad \qquad  n_\mu n^\mu = 1 \,.
\end{equation}
These tangent vectors have unit norm.  the details of the properties of the `world function', $\mu(x, x')$ and its derivatives 
can be obtained in \cite{Synge:1960ueh,doi:10.1063/1.1664961,Allen:1985wd}. 
We also need the one more derivative on these tangent vectors
\begin{equation}
D_\nu n_\sigma = A ( g_{\nu\sigma}  - n_\nu n_\sigma) \,,
\end{equation}
where $A$ is a function of $\mu$ and obeys the differential equation
\begin{eqnarray}
\frac{dA}{d\mu} = - C^2,  &&\qquad \frac{dC}{d\mu} = - AC, \\ \nonumber
 A= \frac{1}{L} \coth\big( \frac{\mu }{L} \big), && \qquad C =  -\frac{1}{L \sinh\big( \frac{\mu}{L} \big) } \,.
\end{eqnarray}
Finally $\Lambda(x, x')$  in (\ref{ansatz}) is the parallel propagator for Dirac spinors in $AdS_2$, which obeys the 
equation 
\begin{equation} \label{parallelpropeq}
D_\nu \Lambda( x, x') = \frac{1}{2} ( A+C) ( \gamma_\nu \gamma^\sigma n_\sigma - n_\nu) \Lambda \,.
\end{equation}
Substituting the ansatz  (\ref{ansatz}) in  the equation for the Green's function (\ref{eqgreenf}), we obtain the 
following equations 
\begin{eqnarray} \label{albeteq}
i \frac{d\hat\alpha}{d\mu}  + \frac{i}{2} ( A + C)\hat\alpha - m \hat\beta = 0, \\ \nonumber
i \frac{d\hat\beta}{d\mu} + \frac{i}{2} ( A- C) \hat\beta + m \hat\alpha = \frac{\delta^2 ( x-x')}{\sqrt{g (x) }} \,.
\end{eqnarray}
Eliminating $\hat\beta$, we  arrive at the second order differential equation for $\hat\alpha$
\begin{equation}\label{eqforalpha}
\frac{d^2 \hat\alpha}{d\mu^2} +  A \frac{d\hat\alpha}{d\mu} - \Big[ \frac{C}{2} ( A+ C) - \frac{1}{4L^2} + m^2  \Big] \hat\alpha= -
m \frac{\delta^2 ( x-x')}{\sqrt{g (x) }} \,.
\end{equation}

To solve the equation (\ref{eqforalpha}), we first observe that the first two terms are just the Laplacian acting on 
$\hat\alpha(\mu)$. 
\begin{eqnarray}
\partial_\nu \hat\alpha( \mu ) &=& n_\nu \frac{d\hat\alpha}{d\mu},  \\ \nonumber
D^\mu D_\mu \hat\alpha( \mu ) &=&  \frac{d^2 \hat\alpha}{d\mu^2} +  A \frac{d\hat\alpha}{d\mu}  \,.
\end{eqnarray}
The remaining terms in (\ref{eqforalpha}) are non-derivative terms and therefore will not affect the leading singularity 
near coincident points that is needed to obtain the delta function on the right hand side of (\ref{eqforalpha}). 
Comparing with the short distance  limit the bosonic Greens function satisfies in (\ref{adsgreen}) and its equation
 (\ref{diffeqg}) \footnote{Note that the differential equation the bosonic Green's function satisfies involves the Laplacian.}, 
we see that we must have 
\begin{equation} \label{behalpha}
\lim_{\xi \rightarrow 1} \hat\alpha( \mu) = -\frac{m}{4\pi} \log( 1- \xi) , \qquad 1-\xi  \sim  \frac{\mu^2}{2 L^2} \,.
\end{equation}
This will fix the normalization for the solution we determine. 
Let us change variable  in the differential equation (\ref{eqforalpha})  to 
\begin{equation} \label{defy}
y = \frac{1}{\cosh^2 \big( \frac{\mu}{2L} \big)} \,.
\end{equation}
Then the equation for $\hat\alpha$ becomes 
\begin{eqnarray}
y^2 ( 1- y) \frac{d^2 \hat\alpha}{dy^2}  - y^2 \frac{d\hat\alpha}{dy}  -\frac{1}{4} ( -1 +  4L^2 m^2  - y) \hat\alpha & =& 
 -m \frac{\delta ( r- r') \delta ( \theta- \theta') }{  \sinh r }\,.  
\end{eqnarray}
Examining the corresponding homogenous equation, we obtain the following solutions 
\begin{equation}
 y^{\frac{1}{2}  + Lm  }  {}_2 F_1( L m, 1+Lm, 1+ 2Lm, y) , \qquad
  y^{\frac{1}{2}  - Lm  }  {}_2 F_1(- L m, 1-Lm, 1- 2Lm, y) \,.
\end{equation}
It is clear that the normalizable solution, that is the solution which is well behaved at infinity is given by 
\begin{equation}
\hat\alpha (y) = K  y^{\frac{1}{2}  + L|m|  }  {}_2 F_1( L| m|, 1+L|m|, 1+ 2L|m|, y)\,.
\end{equation}
We can fix the constant $K$ by examining the behaviour of the hypergeometric function at 
 the origin $y\rightarrow 1$, which is given by 
\begin{equation} \label{yonelimi}
\lim_{y\rightarrow 1} \hat\alpha(y) = -K \frac{\Gamma( 1+ 2L|m|) }{ \Gamma(L|m|) \Gamma( 1+L|m|) } 
\left( \log( 1-y) + 2 \gamma + \psi( L|m|)  +\psi( 1+ L|m|)   + \cdots \right)\,.
\end{equation}
Here the $\cdots$ refer to terms that are suppressed at least as $(1-y) \log ( 1-y) $. 
From the definition of $y$ in (\ref{defy}), we see that 
\begin{equation}\label{yonelimi1}
\lim_{y\rightarrow 1} 1-y= \frac{\mu^2}{4L^2} =  \frac{1-\xi}{2}  \,.
\end{equation}
To obtain the relation with $\xi$, we have used the equation (\ref{defmuxi}). 
As we have discussed earlier,  the leading behaviour if $\hat\alpha$  is fixed  by (\ref{behalpha}).
From (\ref{yonelimi}) and (\ref{yonelimi1}), this implies 
\begin{equation}
K = \frac{m}{4\pi} \frac{ \Gamma ( 1+ L|m|) \Gamma( L|m|)}{ \Gamma( 1 + 2 L|m| ) } \,,
\end{equation}
therefore
\begin{equation} \label{solutionfora}
\hat\alpha ( \mu ) =  \frac{m}{4\pi} \frac{ \Gamma ( 1+ L|m|) \Gamma( L|m|)}{ \Gamma( 1 + 2 L|m| ) }  y^{\frac{1}{2}  + L|m|  } 
 {}_2 F_1( L| m|, 1+L|m|, 1+ 2L|m|, y) \,.
\end{equation}
We can find $\hat\beta(\mu)$ using the first equation of (\ref{albeteq}). 
Using this input, we obtain the following behaviour for $\hat\alpha$ and  $\hat\beta$ at the coincident point
\begin{eqnarray}  \label{limitab}
\lim_{\mu \rightarrow 0} \hat\alpha(\mu) &=&- \frac{m}{4\pi} \Big[ 
\log\big( \frac{\mu^2}{4L^2} \big) + 2 \gamma + \psi( L|m|)  +\psi( 1+ L|m|) +  O(\mu^2 \log \mu )   \Big], \\ \nonumber
\lim_{\mu \rightarrow 0 } \hat\beta( \mu ) &=& -\frac{i}{\sqrt{2} \pi \mu}   \Big[ 1 +  O( \mu^2 \log \mu) \Big] \,.
\end{eqnarray}

Now we can put all the ingredients together to obtain the behaviour of the 
Greens function in (\ref{ansatz}). 
Firstly the parallel propagator for the Dirac spinor obeys the equation given in (\ref{parallelpropeq}). 
From this it is easy to see that as $\mu \rightarrow 0$, it admits a solution of the form
\begin{equation}\label{explamb}
\lim_{\mu\rightarrow 0} \Lambda(x, x') = 1- \frac{1}{8} \mu(x, x')  \omega_\nu^{ab} [\gamma_a, \gamma_b] n^\nu + O(\mu^2) \,.
\end{equation}
What is important to realise from this expansion near $\mu=0$ is that $\Lambda(x, x')$ is independent 
of the mass $m$. Consider the 
Green's function
\begin{equation} 
G(x, x') = \left[ \hat\alpha (\mu ) \gamma_t + \hat\beta(\mu) n_\nu \gamma^\nu ) \right] \Lambda( x, x')\,.
\end{equation}
We see that from  relation of $\hat\beta$ in terms of $\hat\alpha$ in (\ref{albeteq}) and the solution of $\hat\alpha$ in (\ref{solutionfora}), 
that $\hat\beta$ is a function of $|m|$. 
Therefore the term $\hat\beta n_\mu \gamma^\mu \Lambda(x, x')$ will not be discontinuous at $m=0$. 
Furthermore, the finite part of this  term cannot be discontinuous in its first derivative. 
This is because   the leading singularity in 
$\hat\beta$ is a pole in $\mu$ (\ref{limitab}), but whose coefficient is independent of the mass $m$. 
The pole needs to be subtracted, a finite term can arise from this pole multiplying the linear term in $\mu$ 
in the expansion of $\Lambda$ given in (\ref{explamb}) or from $n_\nu$. 
But both $\Lambda$ and $nu_\mu$ are independent of the mass and therefore the finite term arising from the 
second term $\hat\alpha$ in the Green's function is independent of $m$ and therefore continuous at $m=0$. 

Now let us examine the term $\hat\alpha(\mu) \gamma_t \Lambda(x, x')$ in the Green's function. 
The leading singularity in $\hat\alpha$ as $\mu\rightarrow 1$ is proportional to $\log(\mu)$, this is multiplied 
by $\Lambda (x, x')$. This singularity should be removed by the point split regularization, the finite term is  therefore
obtained by looking at the finite term of both $\hat\alpha$ and $\Lambda$ in the $\mu\rightarrow 0$ limit. 
Therefore we can set $\Lambda =1$ and extract the finite term from  the expansion of $\hat\alpha $ given in 
(\ref{limitab}).  This leads to  
\begin{eqnarray} \label{expfermibi}
\langle \bar\psi\gamma_t\psi \rangle&=& 
\left. 
-\lim_{\mu \rightarrow 0} {\rm Tr} ( \gamma_t G(x, x) )\right|_{\rm regularised}, \\ \nonumber
   &=& \frac{2m}{4\pi} \Big[ - \log( 4)  + 2 \gamma + \psi( L|m|)  +\psi( 1+ L|m|) \Big]\,,
\end{eqnarray}
where we have regularized by subtracting the UV divergence which is proportional to $\log\big( \frac{\mu^2}{L^2}) $
\footnote{The negative sign in the first line of (\ref{expfermibi}) is because the Greens function is defined as 
 $G_{\alpha\beta}(x, x') = \langle \psi_\alpha (x)  \bar \psi_{\beta} (x') \rangle$.}.  
Of course this definition of the expectation value of the fermion bilinear is ambiguous up to a constant. 
Let us examine the behaviour of  the following terms as $m\rightarrow 0$
\begin{equation}
\lim_{m\rightarrow0} (  2 \gamma + \psi( L|m|)  +\psi( 1+ L|m|) ) = -\frac{1}{L|m|} + O ( L|m|) \,.
\end{equation}
Substituting this behaviour in (\ref{expfermibi}), it is clear that the expectation value of the fermion bilinear  is discontinuous 
at $x=0$ and the jump in discontinuity is given by 
\begin{equation} \label{jumpa}
\pi L \left. \Big( \langle \bar \psi \gamma_t \psi\rangle|_{m\rightarrow 0^+ } - \langle \bar \psi \gamma_t \psi\rangle|_{m\rightarrow 0^- }
\Big)\right| = -1\,.
\end{equation}
We see that this agrees with the result obtained by the numerical evaluation of the partition function in (\ref{expectjump}). 
This discontinuity is not affected by the ambiguity of the constant in the definition of the expectation value in (\ref{expfermibi}). 
Therefore we conclude that the fermion bilinear is $\langle \psi\gamma_t\psi \rangle$ evaluated from the partition function corresponding to the 
action in (\ref{act2}) is discontinuous whenever the mass of the fermion crosses the BF bound.

If one proceeds on the similar lines to evaluate the expectation value of $\bar\psi\psi$ for the action $S_I$ given in (\ref{act1}) we obtain the same discontinuity. Also note from (\ref{jumpa}) the jump   is given by  $\frac{1}{\pi L}$ which vanishes in the flat space limit. 

\subsubsection*{Fermion bilinear on $\mathbb{R}^2$}

Here we evaluate the expectation of the fermion bilinear when the theory is considered on $\mathbb{R}^2$.
From the action (\ref{act2}), we see that the coincident limit of the fermion bilinear is given by 
\begin{equation}
\langle \bar\psi (\vec y)  \gamma_t \psi (0)  \rangle_{\mathbb{R}^2} =
- \frac{2}{ 4\pi^2} \int_{-\Lambda}^\Lambda  dp_1 dp_2 \frac{(  p_1 - m) e^{ i \vec p \cdot \vec y } }{ ( p_1 -m)^2  + p_2^2} 
\end{equation}
Here we have introduced a cutoff to regulate the integral, the factor of 2 in the numerator is due to the taking the trace over the 
$\gamma$ matrices. 
The coincident limit is given by 
\begin{equation}
\langle \bar\psi   \gamma_t \psi   \rangle_{\mathbb{R}^2} =
- \frac{2}{ 4\pi^2} \int_{-\Lambda}^\Lambda  dp_1 dp_2 \frac{(  p_1 - m)  }{ ( p_1 -m)^2  + p_2^2} 
\end{equation}
It is convenient to perform the $p_2$ integral first \footnote{The result is invariant if the $p_1$ integral is performed first.}.
This results in 
\begin{eqnarray}
\langle \bar\psi  \gamma_t \psi  \rangle_{\mathbb{R}^2} &=& -\frac{1}{2\pi} \int_{-\Lambda}^{\Lambda} dp_1 \frac{p_1 -m}{ |p_1-m| }\, ,  \nonumber \\
&=& \frac{m}{\pi}\, .
\end{eqnarray}
The important point to note in this result is not the value of the expectation value of the fermion bilinear which is subject to 
the definition of the regulator, but that it is continuous in $m$ which is the only scale in the theory. 
Therefore the expectation value of the fermion bilinear  on $\mathbb{R}^2$ 
does not exhibit any jumps as seen in the case of $AdS_2$ in (\ref{jumpa}).

We have performed the same analysis for the expectation value of the fermion bilinear 
$\langle \bar\psi \psi\rangle_{\mathbb{R}^2}$
with the action (\ref{act2}). 
The result is the same, there is no discontinuity in the expectation value at $m=0$ contrary to what 
is seen for the case of $AdS_2$ where the discontinuity is determined by the inverse radius of $AdS_2$.

 \section{Supersymmetric actions on $AdS_2\times S^1$ and $AdS_2\times S^2$ } \label{localactions}

 In this section we demonstrate that
  the supersymmetric  actions one obtains by considering matter multiplets on $AdS_2\times S^1$  or $AdS_2\times S^2$
  are such that 
  the Kaluza-Klein masses  on $S^1$ or $S^2$ can be dialled so that they saturate the BF bound. 
  This implies,  by the discussion in sections \ref{boson}, \ref{fermion}, that the partition function of these theories are 
  not smooth at the BF bound.  Moreover expectation value of the  boson bilinear have a kink and the 
  fermion bilinear expectation 
  value is  discontinuous whenever parameters are dialled so that Kaluza-Klein masses cross the BF bound.

  \subsection{Chiral multiplet on $ AdS_{2}\times S^{1}$} \label{partads2s1}
Our first example would be the supersymmetric theory of the  free chiral multiplet on the 
$AdS_{2}\times S^{1}$. 
This has been studied in \cite{David:2018pex},  where the partition function of the theory in the background of 
the vector multiplet was evaluated using the Greens function method. 
We revisit the example below but will set the vector multiplet background to zero. 
To connect with the discussion in the previous sections, we consider the metric of the space to be
\be
ds^{2}=U^{2}d\tau^{2}+L^{2}(dr^{2}+\sinh^{2}r\,d\theta^{2})\,,
\ee
where $\tau\in[0,2\pi)$.
The metric background together with the auxiliary fields 
\be \label{susyback}
A_{\tau}=V_{\tau}=\frac{U}{L}
\ee
admits solutions of the killing spinor equations. For more details about the notations and the killing spinor solution, we refer to appendix~\ref{appen}.

The supersymmetric action of free chiral multiplet on the above background is
\bea \label{susyact1}
S&=&\int\,d^{3}x\sqrt{g}\Big[\mathcal D_{\mu}\bar\phi \mathcal D_{\mu}\phi+\Big(-\frac{\Delta}{4}R+\frac{1}{2}\left(\Delta-\frac{1}{2}\right)V^{2}\Big)\bar\phi\phi+\bar\psi\slashed {\mathcal D}\psi\Big]\,,
\eea
where
\bea
&&\mathcal D_{\mu}\phi=\p_{\m}\phi-\frac{i}{2}(\Delta-1)V_{\m}\phi\,,\nn\\
&&{\mathcal D}_{\mu}\psi=\nabla_{\mu}\psi-\frac{i}{2}\Delta V_{\mu}\psi\,,
\eea
where $\Delta$ is the R-charge of the scalar field.
The Ricci scalar on $AdS_2$ is given by 
\begin{equation}
R = \frac{2}{L^2}
\end{equation}
Here we have incorporated the negative sign of the curvature in the curvature coupling of the action in (\ref{susyact1}). 
To see this we can set the background $V=0$ and choose the R-charge $\Delta =\frac{1}{2}$ and observe that 
the action for the scalar reduces to the conformal coupled scalar on $AdS_2 \times S^1$. 
Let us now proceed  by substituting the  supersymmetric background for $V_{\mu}$ in (\ref{susyact1}) 
and the Ricci scalar on $AdS_2$, we obtain
\be \label{susyact2}
S=\int\,d^{3}x\sqrt{g}\Big[\mathcal D_{\mu}\bar\phi \mathcal D_{\mu}\phi-\frac{1}{4L^{2}}\bar\phi\phi+\bar\psi\slashed {\mathcal D}\psi\Big]\,.
\ee
Next, we reduce the above action on S$^{1}$ to obtain the action on AdS$_{2}$. 

\subsubsection*{Bosons}
We start with the scalar field.  The expansion of the scalar in term of its Fourier modes is  
\be
\phi(\tau,r,\theta)=\sum_{n}e^{in\tau}\phi_{n}(r,\theta)\,.
\ee
Substituting this expansion in (\ref{susyact2}), 
the action for the  scalar field  becomes 
\bea \label{mbosact}
S &=&\int\,d^{3}x\sqrt{g}\Big[\mathcal D_{\mu}\bar\phi \mathcal D_{\mu}\phi-\frac{1}{4L^{2}}\bar\phi\phi\Big]\,, \\ \nonumber
&=&2\pi U\sum_{n -\infty}^\infty
\int d^{2}x\,\sqrt{\tilde g}\left[\tilde g^{ij}\p_{i}\bar\phi_{n}\p_{j}\phi_{n}+\Big(  \big(\frac{n}{U} -\frac{\Delta-1}{2L} \big)^{2}-\frac{1}{4L^{2}}\Big)|\phi_{n}|^{2}\right]\,.
\eea
In the above $\tilde g_{ij}$ is the metric on AdS$_{2}$ of radius $L$. 
Therefore the mass of the $n$-th Kaluza-Klein mode is  
\be
m_{n}^{2}=-\frac{1}{4L^{2}}+\frac{x_{n}^{2}}{L^{2}},\quad\text{where}\quad x_{n}^{2}=\Big( \frac{ nL}{U} -\frac{(\Delta-1)}{2}\Big)^{2}\,.
\ee
From the discussion in section  \ref{boson}, we see that the partition function is not smooth whenever $x_n =0$ or the 
Kaluza-Klein mass saturates the BF bound. 
For a given R-charge, these values are determined by the ratio of the radii  of $AdS_2$ to $S^1$, i.e., whenever  the  ratio is such that 
\begin{equation}\label{cond1}
\frac{U(\Delta -1)}{2 L} \in \mathbb{Z}
\end{equation}
there exists a Kaluza-Klein mass which saturates the BF bound, and therefore the partition function is not smooth
as the ratio $\frac{L}{U}$ is varied. 
Furthermore, the expectation value
\begin{equation}
\langle \phi( \tau, r, \theta) \bar{\phi}( \tau, r, \theta) \rangle = \sum_{n=-\infty}^\infty \langle |\phi_n( r, \theta)|^2 \rangle , 
\end{equation}
shows a kink whenever as ratio $\frac{L}{U}$ is varied across points which satisfy (\ref{cond1}).

\subsubsection*{Fermions}
Similarly,  let us expand the fermions in terms of Fourier modes along the $S^1$ direction
\begin{equation}
\psi (\tau, r, \theta) = \sum_{n} e^{ i ( n+ r_0)  \tau } \psi_n(r, \tau) , \qquad r_0 = 0, \;{\mbox or}\;\; \frac{1}{2} \,.
\end{equation}
Here, $r_0$ determines the periodicity of the fermion, for $r_0=0, \frac{1}{2}$, the fermions are periodic  or anti-periodic, respectively. 
Substituting this expansion in the fermionic part of the action~\eqref{susyact2}, we obtain 
\be
\int d^{3}x\,\sqrt{g}\,\wt\psi\slashed {\mathcal D}\psi=2\pi U 
\sum_{n}\int d^{2}x\,\sqrt{\tilde g}\left[\bar\psi_{n}\slashed {\nabla}\psi_{n}+\frac{i}{L}\big(n-\frac{U\Delta}{2L}+r_{0}\big)\bar\psi_{n}\g_{1}\psi_{n}\right]\,.
\ee
Here $\slashed {\nabla}$ is the covariant Dirac operator on AdS$_{2}$. 
Comparing with the discussion in the section \ref{fermion}, we see that
\be
x_{n}=-\Big((n+r_{0})-\frac{U\Delta}{2 L}\Big)\,.
\ee
Therefore, whenever the ratio of the radii, $\frac{L}{U}$, is such that 
\begin{equation} \label{cond2}
\frac{U\Delta}{2L} \in \mathbb{Z} + r_0\,,
\end{equation}
the partition function for the fermions has a kink. Furthermore, the expectation value of the fermion bilinear 
is given by 
\begin{equation}
\langle\bar \psi (\tau, r, \theta )\gamma_1 \psi(\tau, r, \theta) \rangle = 
\sum_{n=-\infty}^\infty \langle \bar\psi_n (r, \theta) \g_{1}\psi_n(r, \theta) \rangle \,.
\end{equation}
Again from the discussion in section (\ref{fermion}), we see that this expectation value has a discontinuity whenever
the ratio of the radii satisfies (\ref{cond2}). 

Finally we remark that the discontinuity for the fermion in (\ref{cond2}) occurs at different values of the ratio
$\frac{L}{U}$ compared to that of the boson in (\ref{cond1}).

\subsection{Hypermultiplet on AdS$_{2}\times$S$^{2}$} \label{partads2s2}

Our next example is the action of a free hypermultiplet on AdS$_{2}\times$S$^{2}$. 
We consider a supersymmetric background which admits one or more killing spinor and it consists of the background metric,
\be\label{ads2s2}
ds^{2}=L^{2}(dr^{2}+\sinh^{2}r\,d\theta^{2})+U^{2}(d\varphi^{2}+\sin^{2}\varphi\,d\psi^{2})\,,
\ee
together with auxiliary graviphoton field $T_{\m\n}$ and the scalar field $M$. These auxiliary fields are given by
\begin{eqnarray} \label{background}
&& M=\frac{1}{L^{2}}-\frac{1}{U^{2}},\quad T^{ab}\g_{ab}=-i\alpha\,\mathbb{I}\otimes\tau_{3}\,, 
\end{eqnarray}
where $\alpha =\Big(\frac{1}{4L}+\frac{1}{4U}\Big)$\,. We follow~\cite{Hama:2012bg} for the action of hypermultiplet and backgrounds fields. 
The Ricci scalar of the metric in (\ref{ads2s2})  is given by $R=\frac{2}{L^{2}}-\frac{2}{U^{2}}$, here again we are using 
the positive sign for the curvature on $AdS_2$ and negative sign for the curvature on $S^2$.

To demonstrate that the partition function of the hypermultiplet is not smooth, we also need a background vector multiplet. 
The vector multiplet is non-dynamical, and we choose the BPS background value, which is obtained by solving the BPS equations $\delta\lambda_{i}=0$. The variation $\delta\lambda_{i}$ is determined by the Killing spinors on $AdS_2$ which is given in the appendix~\ref{appen}. The BPS solution is
\be
a_{\m}=\hat{m}(1-\cos\varphi)\,d\psi\,,\quad D_{ij}=0,\quad\sigma=-\frac{\hat{m}}{2U},\quad\bar\sigma=\frac{\hat{m} }{2U}\,.
\ee
Here $\hat{m}$ is the magnetic flux through $S^{2}$, $D_{ij}$ is the auxillary field and $\sigma$ is the scalar in the  ${\cal N}=2$ vector multiplet. 
The action for the free hypermultiplet is
\bea \label{laghyper}
\mathcal L&=&D_{\m}\phi_{1}^{\dagger}D_{\m}\phi_{1}+D_{\m}\phi_{2}^{\dagger}D_{\m}\phi_{2}+\Big[\frac{1}{4}(-R+M)-4g^{2}\sigma\bar\sigma\Big](|\phi_{1}|^{2}+|\phi_{2}|^{2})\nn\\
&&-\frac{i}{2}\overline{\hat\psi}\g^{\m} D_{\m}\psi-i\overline{\hat\psi}\g^{\m\n}\psi T_{\m\n}+g\overline{\hat\psi}(\sigma P_{+}+\bar\sigma P_{-})\psi\,.
\eea
Here $\phi_{1}$ and $\phi_{2}$ are two complex scalar fields of the hypermultiplet.
The fields $\hat\psi$ and $\psi$ are two 4-component Dirac spinors satisfying the reality property
\be \label{constraintsp}
\overline{\hat\psi}=\psi^{\dagger},\quad \hat\psi^{\dagger}=-\overline\psi\,.
\ee
Also, $P_{\pm}$ are projection operators that project a 4-component fermion to its positive and negative chiral part, respectively
\footnote{We have combined the 2 Weyl spinors in the action of \cite{Hama:2012bg}  to a 
Dirac spinor  together with the reality constraint 
in (\ref{constraintsp}).}.
Furthermore, the covariant derivatives are
\bea
&&D_{\m}\phi_{1}=(\p_{\m}-iga_{\m})\phi_{1},\quad D_{\m}\phi_{2}=(\p_{\m}+iga_{\m})\phi_{2}\,,\nn\\
&&D_{\m}\psi=(\nabla_{\m}-iga_{\m})\psi,\quad D_{\m}\hat\psi=(\nabla_{\m}+iga_{\m})\hat\psi\,.
\eea
Here $a_{\m}$,  is the background $U(1)$ gauge field which  couples to $Sp(1)$ index. 
We have considered only a single hypermultiplet. In principle one could consider more hypers, say $r$ hypers. In that case 
this background should be thought of as turning on the gauge field in   the Cartan for of $Sp(r)$.

\subsubsection*{Bosons}

Next, we proceed as previously. Firstly, we start with the scalar fields. We discuss only one scalar field since both scalar fields have similar kinetic terms.
We expand the scalar field in Monopole harmonics as
\begin{equation}
\phi_1(r, \theta, \varphi, \psi) = \sum_{\ell, p } \phi_{1 g\hat m; \ell, p }(r, \theta)  Y_{g\hat m, \ell, p} (\varphi, \psi) 
\end{equation}
These harmonics satisfy the  following eigen value equation on $S^2$. 
\begin{eqnarray}
&&- \frac{1}{U^2 \sin^2 \varphi} \left[ \sin\varphi \frac{\partial}{\partial\varphi} \Big(  \sin\varphi  \frac{\partial}{\partial\varphi} \Big)
+ \Big(\frac{\partial}{\partial\psi}  - i g\hat m ( 1- \cos\psi )  \Big)^2  \right]
Y_{g\hat m,\ell,p}(\varphi,\psi)  \nonumber \\
&& \qquad\qquad\qquad=\frac{\ell(\ell+1)-g^{2}\hat m^{2}}{U^{2}}Y_{gm,\ell,p}(\varphi,\psi)\,,
\end{eqnarray}
with $\ell\geq |g \hat m|$. 
Substituting this expansion in the action for the scalar $\phi_1$, we obtain 
\bea
\mathcal L^{(1)}_{s}&=&-g^{\m\n}\phi_{1}^{\dagger}D_{\m}D_{\n}\phi_{1}+\Big[\frac{1}{4}(-R+M)-4g^{2}\sigma\bar\sigma\Big]|\phi_{1}|^{2}\nn\\
&=&-\phi_{1;g\hat m,\ell,n}^{\dagger}\Box_{AdS_{2}}\phi_{1;g\hat m,\ell,p}+\frac{1}{U^{2}}\Big(\ell(\ell+1)+\frac{1}{4}(1-\frac{U^{2}}{L^{2}})\Big)\phi_{1;g\hat m,\ell,p}^{\dagger}\phi_{1;g\hat m,\ell,p}
\eea
where $\Box_{AdS_{2}}$ is the Laplacian on AdS$_{2}$, the sum over  $\ell, p $ is implied in the second line. 
 Note that the mass of the $\ell$-the harmonic is given by 
\be
m^{2}_{\ell}=\frac{4\ell(\ell+1)+1}{U^{2}}-\frac{1}{4L^{2}}\,.
\ee
Since $\ell\geq |g\hat m|$, the above mass can not saturate the BF bound. 
Thus, the masses of the  bosons in the hypermultiplet on the supersymmetric background of $AdS_2\times S^2$ do
not saturate the BF bound and therefore, we do not encounter the discontinuity in the partition function which was pointed out in 
section (\ref{boson}). 

\subsubsection*{Fermions}

Next, we look at the action of fermions.  From (\ref{laghyper}) we see that, 
to evaluate the one loop determinant, we need to know the eigen values of the operator $\mathbb{D}$ where
\be
\mathbb{D}\psi=\slashed D\psi+2\g^{\m\n}T_{\m\n}\psi+2ig(\sigma P_{+}+\bar\sigma P_{-})\psi\,.
\ee
Using the convention of the gamma matrices given in the appendix~\ref{appen}, the Dirac operator appearing in the above is sum of the Dirac operator on AdS$_{2}$ and S$^{2}$ as 
\be
\g^{\m}D_{\m}=\mathbb{I}\otimes \slashed D_{AdS_{2}}+\slashed D_{S^{2}}\otimes \tau_{3}\,,
\ee 
where
\be
\slashed D_{AdS_{2}}=\frac{1}{L}\Big(\frac{1}{\sinh\,r}\tau_{1}\p_{\theta}+\tau_{2}\p_{r}+\frac{\tau_{2}}{2}\coth\, r\Big)\,,
\ee
and
\be
\slashed D_{S^{2}}=\frac{1}{U}\Big(\sigma_{2}\p_{\varphi}+\frac{\sigma_{1}}{\sin\varphi}(\p_{\psi}-ig a_{\psi})+\frac{\sigma_{2}}{2}\cot\varphi\Big)\,.
\ee
Thus, the complete fermionic operator whose eigen values needs to be determined is given by 
\be
\mathbb{D}\psi=\psi_{1}\otimes \slashed D_{AdS_{2}}\psi_{2}+\slashed D_{S^{2}}\psi_{1}\otimes \tau_{3}\psi_{2}-2i\alpha\,\psi_{1}\otimes\tau_{3}\psi_{2}+\frac{ig \hat m}{U}\,\sigma_{3}\psi_{1}\otimes\tau_{3}\psi_{2}\,.
\ee
Here $\psi_{1}$ and $\psi_{2}$ are the two component spinors on S$^{2}$ and AdS$_{2}$, respectively.

Next, we want to compute the eigen values of the operator $\mathbb{D}$. 
Given the eigen functions and eign values of the Dirac operator on S$^{2}$ and AdS$_{2}$~\footnote{See the appendix~\ref{appen} for the details of the eigen functions and eigen values in the presence of the magnetic flux.}, i.e.
\be
\slashed D_{AdS_{2}}\psi_{2}=i\zeta_{1}\psi_{2},\quad \slashed D_{S^{2}}\psi_{1}=i\zeta_{2}\psi_{1}\,,
\ee
where $\zeta_{1}=\frac{\lambda}{L}$ with $\lambda\in\mathbb{R}_{+}$ and $\zeta_{2}^{2}=\frac{(\ell+1)^{2}-g^{2}m^{2}}{U^{2}}$, it is easy to see the following relations:
\bea
&&\mathbb{D}(\psi_{1}\otimes\psi_{2})=i\zeta_{1}\,\psi_{1}\otimes\psi_{2}+i\zeta_{2}\,\psi_{1}\otimes\tau_{3}\psi_{2}-2i\alpha\,\psi_{1}\otimes\tau_{3}\psi_{2}+\frac{ig \hat m}{U}\,\sigma_{3}\psi_{1}\otimes\tau_{3}\psi_{2}\,,\nn\\
&&\mathbb{D}(\psi_{1}\otimes\tau_{3}\psi_{2})=-i\zeta_{1}\,\psi_{1}\otimes\tau_{3}\psi_{2}+i\zeta_{2}\,\psi_{1}\otimes\psi_{2}-2i\alpha\,\psi_{1}\otimes\psi_{2}+\frac{ig \hat m}{U}\,\sigma_{3}\psi_{1}\otimes\psi_{2}\,,\nn\\
&&\mathbb{D}(\sigma_{3}\psi_{1}\otimes\psi_{2})=i\zeta_{1}\,\sigma_{3}\psi_{1}\otimes\psi_{2}-i\zeta_{2}\,\sigma_{3}\psi_{1}\otimes\tau_{3}\psi_{2}-2i\alpha\,\sigma_{3}\psi_{1}\otimes\tau_{3}\psi_{2}+\frac{ig \hat m}{U}\,\psi_{1}\otimes\tau_{3}\psi_{2}\,,\nn\\
&&\mathbb{D}(\sigma_{3}\psi_{1}\otimes\tau_{3}\psi_{2})=-i\zeta_{1}\,\sigma_{3}\psi_{1}\otimes\tau_{3}\psi_{2}-i\zeta_{2}\,\sigma_{3}\psi_{1}\otimes\psi_{2}-2i\alpha\,\sigma_{3}\psi_{1}\otimes\psi_{2}+\frac{ig \hat m}{U}\,\psi_{1}\otimes\psi_{2}\,. \nn\\
\eea
Therefore, the eigen values of the operator $\mathbb{D}$ are obtained by finding the eigen values of the matrix
\be
\begin{pmatrix}i\zeta_{1}&i\zeta_{2}-2i\alpha&0&\frac{ig \hat m}{U}\\i\zeta_{2}-2i\alpha&-i\zeta_{1}&\frac{ig \hat m}{U}&0\\0&\frac{ig\hat m}{U}&i\zeta_{1}&-i\zeta_{2}-2i\alpha\\\frac{ig\hat m}{U}&0&-i\zeta_{2}-2i\alpha&-i\zeta_{1}\end{pmatrix}\,.
\ee
The eigen values are
\bea
\pm i\sqrt{\frac{\lambda^{2}}{L^{2}}+\frac{1}{U^{2}}\Big(\ell+\frac{1}{2}-\frac{U}{2L}\Big)^{2}},\quad \pm i\sqrt{\frac{\lambda^{2}}{L^{2}}+\frac{1}{U^{2}}\Big(\ell+\frac{3}{2}+\frac{U}{2L}\Big)^{2}}\,,
\eea
where we have used the explicit form for the eigen values $\zeta_{1,2}$ and the background value of 
$\alpha$ given in (\ref{background}).  Thus, the partition function is
\be
\ln Z=4 \sum_{\ell=|gm|}^{\infty}(\ell+1)\int d\lambda\,\lambda\,\coth\pi\lambda\Big[\ln\Big(\frac{\lambda^{2}}{L^{2}}+\frac{1}{U^{2}}\Big(\ell+\frac{1}{2}-\frac{U}{2L}\Big)^{2}\Big)+\ln\Big(\frac{\lambda^{2}}{L^{2}}+\frac{1}{U^{2}}\Big(\ell+\frac{3}{2}+\frac{U}{2L}\Big)^{2}\Big)\Big]
\ee
Comparing with the discussion presented in the section~\ref{fermion}, we see that the above partition function can be thought of as the partition function of Kaluza-Klein towers of fermions on AdS$_{2}$ with masses
\be
m^{+}_{\ell}=\ell+\frac{3}{2}+\frac{U}{2L},\quad m^{-}_{\ell}=\ell+\frac{1}{2}-\frac{U}{2L}\,.
\ee
In particular, we see that there is a value of the  ratio of the 
radii of $S^2$ to $AdS_2$,  $\frac{U}{L}$ for which the mass $m^{-}_{\ell}$ vanishes for some $\ell$. This happens when
\be\label{FermionCrossratio}
\frac{U}{L}=2\ell+1\,.
\ee
As we have discussed in the section~\ref{fermion}, the partition function will have a kink  whenever the ratio
satisfies the equation (\ref{FermionCrossratio}). 
The expectation value of the fermion bilinear will  be discontinuous at these points.


\section{Conclusions}

The  partition functions of free scalars and fermions on $AdS_2$  
are not smooth as a function of their masses.  This feature is also seen for the case of free scalars or fermions 
on $\mathbb{R}^2$. 
However what is distinct in the case of $AdS_2$ is that the discontinuities seen in the 
expectation values of observables such as the scalar and fermion bilinears are determined by the 
inverse radius of $AdS_2$. 
The most surprising behaviour is that of the  expectation value of fermion bilinear which has a jump as the mass crosses 
the BF bound.  The value of the jump is $-\frac{1}{\pi L}$, where $L$ is the radius of $AdS_2$. 

We have shown that this anomalous behaviour of free theories on $AdS_2$ is exhbited in  the simplest 
supersymmetric theories which involve $AdS_2$ along with compact directions. 
By considering  supersymmetric actions, the chiral multiplet on $AdS_2\times S^1$ and the hypermultiplet on 
$AdS_2 \times S^2$ we have seen these theories also exhibit the same behaviour. 
Here the supersymmetric backgrounds are such that the ratio of the radii of $AdS_2$ and the compact space
plays the role of the parameter which can dial the mass. We see that discontinuities occur for each Kaluza-Klein mode. 

An obvious question is whether the models in ${\cal N}=2$ supergravity admitting  black hole solutions with 
different radii of $AdS_2$ and $S^2$ exhibit this phenomenon. See \cite{Zaffaroni:2019dhb} for a review of black hole solutions 
and \cite{Cassani:2012pj} and \cite{Halmagyi:2013sla} for more specific models. 
Therefore it is important to derive the quadratic action of the fluctuations in the near horizon geometry of these solutions and study the behaviour of the mass spectrum. 

In this paper we have focussed on $AdS_2$, but the phenomenon seen here regarding discontinuities in the behaviour 
of free bosonic or fermionic theories should be true in higher dimensional anti-deSitter spaces. 
This will be interesting to study in detail  further. 

We have seen that key reason for the discontinuity is the fact that as the mass is dialled which 
of wave functions  are normalisable in $AdS_2$ change.  It will be important to 
examine models with interactions   to see if the observation 
found in this paper 
persists.  Since the discontinuity is due to the change in behaviour of the wave functions at infinity, models with interactions which 
do not modify this behaviour will continue to exhibit such discontinuities. 
Indeed we find that the model studied in \cite{David:2018pex} of the chiral multiplet coupled to a background vector also exhibits this behaviour.
It will be interesting to  investigate more models of supersymmetric field theories on 
$AdS_2$ together with a compact space to not only see if discontinuities in fermion bilinears persist but
more importantly study its implications.

\appendix
\section{Notations and Conventions} \label{appen}

\subsection*{Conventions for $AdS_2\times S^1$}
Here we provide the conventions for the gamma matrices and the killing spinors solution on $AdS_2\times S^2$ used in the section (\ref{partads2s1}). Detailed discussion can be found in~\cite{David:2018pex}.
The covariant derivative of a fermion is given by
\be
\nabla_\mu\psi=\left(\p_\mu+\frac{i}{4}\omega_{\mu\,ab}\varepsilon^{abc}\gamma_c\right)\psi,\qquad \varepsilon^{123}=1.
\ee
Our choice for 3-dimensional gamma matrices are
\be
\gamma^1=\begin{pmatrix}1&0\\0& -1\end{pmatrix},\quad \gamma^2=\begin{pmatrix}0&-1\\-1& 0\end{pmatrix},\quad \gamma^3=\begin{pmatrix}0&i\\-i& 0\end{pmatrix}\,.
\ee
They satisfy gamma matrices algebra
\be
\gamma^a\gamma^b=\delta^{ab}+i\varepsilon^{abc}\gamma_c\,.
\ee
These gamma matrices are Hermitian and statisfy
\be
\gamma^{aT}=-C\gamma^aC^{-1},\quad C=\begin{pmatrix}0 & 1\\-1 & 0\end{pmatrix},\quad C^T=-C=C^{-1}\,.
\ee
Solutions of the killing spinor equations are
\be
\epsilon=e^{\frac{i\theta}{2}}\begin{pmatrix}i\cosh\frac{r}{2}\\\sinh\frac{r}{2}\end{pmatrix},\quad \tilde\epsilon=e^{-\frac{i\theta}{2}}\begin{pmatrix}\sinh\frac{r}{2}\\i\cosh\frac{r}{2}\end{pmatrix}\,.
\ee
These spinors generate the killing vector given by
\be
K=\frac{1}{U}\frac{\p}{\p\tau}+\frac{1}{L}\frac{\p}{\p\theta}\,.
\ee
\subsection*{Conventions for $AdS_2\times S^2$} 

Here we list the vielbeins and gamma matrices used for $AdS_2\times S^2$ discussed in section (\ref{partads2s2})
Vielbeins are given as
\be
e^{1}=L\,dr\,,\quad e^{2}=L\,\sinh r\,d\theta\,,\quad e^{3}=U\,d\varphi\,,\quad e^{4}=U\,\sin\varphi\,d\psi\,.
\ee
The gamma matrices are Hermitian matrix $\g^{\m\dagger}=\g^{\m}$.
\be
\g_{1}=\mathbb{I}\otimes \tau_{2},\quad \g_{2}=\mathbb{I}\otimes\tau_{1},\quad \g_{3}=\sigma_{2}\otimes\tau_{3},\quad\g_{4}=\sigma_{1}\otimes\tau_{3}\,.
\ee
Here $\sigma_{i}$ and $\tau_{i}$, for $i=1,2,3$ are Pauli matrices.
The chiral projection operators are given by
\be
P_{\pm}=\frac{1}{2}(1\pm\g_{5}),\quad\text{where}\quad\g_{5}=\g_{1}\g_{2}\g_{3}\g_{4}=-\sigma_{3}\otimes\tau_{3}\,.
\ee
The definition of the $\overline{\psi}$ is
\be
\overline\psi=\psi^{T}C\,,
\ee
where $C$ is the charge conjugation matrix given by
\be
C=-i\sigma_{2}\otimes\tau_{1}\,.
\ee
where $\sigma_{2}$ is a Pauli matrix. With the above charge conjugation matrix, the gamma matrices satisfies
\be
\g^{\m T}=C\g^{\m}C^{-1}\,.
\ee 
The matrix $\g^{\m\n}$ is defined as
\be
\g^{\m\n}=\frac{1}{2}(\g^{\m}\g^{\n}-\g^{\n}\g^{\m})\,.
\ee
The killing spinor equation is given by
\be
D_{\m}\xi^{i}+\g^{\alpha\beta}T_{\alpha\beta}\g_{\m}\xi^{i}=\g_{\m}\eta^{i}\,,
\ee
and the auxiliary field $M$ is determined by the equation
\be
\g^{\m}\g^{\n}D_{\m}D_{\n}\xi^{i}+4\g^{\m}D_{\m}T_{ab}\g^{ab}\xi^{i}=M\xi^{i}\,.
\ee
The spinors $\xi^{i}$ are symplectic Majorana spinors satisfying the reality property
\be
\xi^{i\,\dagger}=\epsilon_{ij}\xi^{j\,T}C\,.
\ee
The solutions are given by
\be
\xi^{1}=\frac{i}{\sqrt{2}}e^{\frac{i}{2}(\theta+\psi)}\begin{pmatrix}\sin\frac{\varphi}{2}\sinh\frac{r}{2}\\\cosh\frac{r}{2}\sin\frac{\varphi}{2}\\\cos\frac{\varphi}{2}\sinh\frac{r}{2}\\\cos\frac{\varphi}{2}\cosh\frac{r}{2}\end{pmatrix},
\quad \xi^{2}=\frac{i}{\sqrt{2}}e^{-\frac{i}{2}(\theta+\psi)}\begin{pmatrix}\cos\frac{\varphi}{2}\cosh\frac{r}{2}\\\sinh\frac{r}{2}\cos\frac{\varphi}{2}\\-\sin\frac{\varphi}{2}\cosh\frac{r}{2}\\-\sin\frac{\varphi}{2}\sinh\frac{r}{2}\end{pmatrix}\,.
\ee
These spinors generate the killing vector field given by
\be
K=\frac{1}{L}\frac{\p}{\p\theta}-\frac{1}{U}\frac{\p}{\p\psi}\,.
\ee
\subsection*{Monopoles scalar harmonics}

The eigen function of the Laplace operator on unit sphere in the presence of the background magnetic flux satisfies the equation
\be
-\Box_{S^{2}}Y_{gm,\ell,p}(\varphi,\psi)=(\ell(\ell+1)-g^{2}m^{2})Y_{gm,\ell,p}(\varphi,\psi)\,,
\ee
where $\ell=|g\hat m|,|g \hat m|+1,...$ and $-\ell\leq p\leq \ell$. The explicit form of the Laplace operator is
\be
\Box_{S^{2}}=\frac{1}{\sin\varphi}\p_{\varphi}( \sin\varphi ) +\frac{1}{\sin^{2}\varphi}(\p_{\psi}-ig \hat m(1-\cos\varphi))^{2}
\ee
and the normalized eigen function (northern hemisphere) is
\begin{eqnarray}
Y_{g\hat m,\ell,p}(\varphi,\psi)&=&e^{i(p+|g\hat m|)\psi}\sqrt{\frac{2\ell+1}{4\pi}}\sqrt{\frac{(\ell-p)!(\ell+p)!}{(\ell+|g\hat m|)!(\ell-|g\hat m|)!}} \\ \nonumber
&& \times \Big(\cos\frac{\varphi}{2}\Big)^{p-|g\hat m|}\Big(\sin\frac{\varphi}{2}\Big)^{p+|g\hat m|}P_{\ell-p}^{(p+|g\hat m|,p-|g\hat m|)}(\cos\varphi)\,.
\end{eqnarray}

\subsection*{Spinor eigen functions on $AdS_2$}

The eigen functions of the Dirac operator on AdS$_{2}$ satisfy
\be
\slashed D_{AdS_{2}}\chi^{\pm}_{p}(\lambda)=\pm i\lambda\,\chi^{\pm}_{p}(\lambda),\quad \slashed D_{AdS_{2}}\eta^{\pm}_{p}(\lambda)=\pm i\lambda\,\eta^{\pm}_{p}(\lambda)\,.
\ee
These are given by
\bea
&&\chi^{\pm}_{p}(\lambda)=\frac{1}{4\pi}\Big|\frac{\Gamma(1+p+i\lambda)}{\Gamma(p+1)\Gamma(\frac{1}{2}+i\lambda)}\Big|e^{i(p+\frac{1}{2})\theta}\times\nn\\
&&\qquad\qquad\begin{pmatrix}\frac{i\lambda}{p+1}\cosh^{p}\frac{r}{2}\sinh^{p+1}\frac{r}{2}\,{}_{2}F_{1}(p+1+i\lambda,p+1-i\lambda;p+2;-\sinh^{2}\frac{r}{2})\\\pm\cosh^{p+1}\frac{r}{2}\sinh^{p}\frac{r}{2}\,{}_{2}F_{1}(p+1+i\lambda,p+1-i\lambda;p+1;-\sinh^{2}\frac{r}{2})\end{pmatrix}\,,\nn\\
&&\eta^{\pm}_{p}(\lambda)=\frac{1}{4\pi }\Big|\frac{\Gamma(1+p+i\lambda)}{\Gamma(p+1)\Gamma(\frac{1}{2}+i\lambda)}\Big|e^{-i(p+\frac{1}{2})\theta}\times \\
&&\qquad\qquad\begin{pmatrix}\cosh^{p+1}\frac{r}{2}\sinh^{p}\frac{r}{2}\,{}_{2}F_{1}(p+1+i\lambda,p+1-i\lambda;p+1;-\sinh^{2}\frac{r}{2})\\\pm\frac{i\lambda}{p+1}\cosh^{p}\frac{r}{2}\sinh^{p+1}\frac{r}{2}\,{}_{2}F_{1}(p+1+i\lambda,p+1-i\lambda;p+2;-\sinh^{2}\frac{r}{2})\end{pmatrix}\,, \nn
\eea
where $p\in\mathbb {Z}$, $0\leq p<\infty$ and $0<\ell<\infty$.

\subsection*{Monopole spinor harmonics}

The eigen functions on $S^{2}$ in the presence of the magnetic flux satisfy
\be
\slashed D_{S^{2}}\chi^{\pm}_{\ell,p}=\pm i\sqrt{(\ell+1)^{2}-g^{2}m^{2}}\,\chi^{\pm}_{\ell,p},\quad \slashed D_{S^{2}}\eta^{\pm}_{\ell,p}=\pm i\sqrt{(\ell+1)^{2}-g^{2}m^{2}}\,\eta^{\pm}_{\ell,p}\,,
\ee
Here $\slashed D_{S^2}$ is given by 
\begin{equation}
\slashed D_{S^2}  = e_i^{\; a} \gamma^i \Big( \partial_a - i g\hat m A_a  + \frac{1}{8} \omega_{a\, ij} [\gamma^i, \gamma^j] \Big) \, ,
\end{equation}
$a \in \{ \varphi, \psi \}$ the coordinates on $S^2$, $i, j$ are the corresponding flat space indices.  
$\gamma^i$ are the Dirac matrices and 
$\omega_{a\,  ij} $ is the spin connection on $S^2$. 
The non-zero value of the
monopole vector potential is given by 
\begin{equation}
A_\psi = \hat m ( 1- \cos \varphi)\, .
\end{equation}
The explicit form of the normalized eigen function on a S$^{2}$ in the northern hemisphere are
\be
\chi^{\pm}_{\ell,p}=\mathcal Ne^{i(p+g\hat m+\frac{1}{2})\psi}\begin{pmatrix}\pm\sqrt{\frac{\ell+1-gm}{\ell+1+g\hat m}}(\cos\frac{\varphi}{2})^{p-g\hat m}(\sin\frac{\varphi}{2})^{1+g\hat m+p}P^{(1+p+g\hat m,p-g\hat m)}_{\ell-p}(\cos\varphi)\\(\cos\frac{\varphi}{2})^{1+p-gm}(\sin\frac{\varphi}{2})^{g\hat m+p}P^{(p+g\hat m,1+p-g\hat m)}_{\ell-p}(\cos\varphi)\end{pmatrix}\,.
\ee
The above is smooth for $\ell\geq |g\hat m|$ and $-g\hat m\leq p\leq \ell$.
\be
\eta^{\pm}_{\ell,p}=\mathcal Ne^{-i(p-g\hat m+\frac{1}{2})\psi}\begin{pmatrix}(\cos\frac{\varphi}{2})^{1+p+gm}(\sin\frac{\varphi}{2})^{p-gm}P^{(p-g\hat m,1+p+g\hat m)}_{\ell-p}(\cos\varphi)\\\mp\sqrt{\frac{\ell+1+g\hat m}{\ell+1-g\hat m}}(\cos\frac{\varphi}{2})^{p+g\hat m}(\sin\frac{\varphi}{2})^{1+p-g\hat m}P^{(1+p-g\hat m,p+g\hat m)}_{\ell-p}(\cos\varphi)\,
\end{pmatrix}\,.
\ee
The above exist for $g\hat m\leq p\leq \ell$ and $\ell\geq |g\hat m|$.
The normalization constant is  given by
\begin{equation}
\mathcal N=\sqrt{\frac{(\ell+1)}{4\pi }\frac{(\ell+p+1)!(\ell-p)!}{(\ell+g\hat m)!(\ell-g\hat m+1)!}}\, .
\end{equation}

\bibliographystyle{JHEP}
\bibliography{references} 

 \end{document}